\documentclass[11pt,a4paper]{article}
\pdfoutput=1

\usepackage{jcappub}

\usepackage{amsmath}
\usepackage{amssymb}
\usepackage{bm}
\usepackage{booktabs}
\usepackage{siunitx}
\usepackage{microtype}
\usepackage{placeins}

\graphicspath{{figures/}}

\hypersetup{
  colorlinks=true,
  linkcolor=blue,
  citecolor=blue,
  urlcolor=blue
}

\newcommand{\dd}{\mathrm{d}}
\newcommand{\thv}{\bm{\theta}}
\newcommand{\qphi}{q_{\phi}}
\newcommand{\pinj}{\pi_{\mathrm{inj}}}
\newcommand{\pdet}{p_{\mathrm{det}}}
\newcommand{\Stilde}{\widetilde{\mathcal S}}
\newcommand{\DeltaIS}{\Delta I_{\Stilde}}

\DeclareMathOperator{\supp}{supp}

\newcommand{\safeinput}[1]{%
  \IfFileExists{#1.tex}{%
    \input{#1}%
  }{%
    \typeout{*** WARNING: `#1.tex' not found; skipping this file. ***}%
  }%
}

\newcommand{\licomment}[1]{}
\newcommand{\sm}[1]{}

\title{\boldmath Hierarchical Population Inference with Normalizing Flows for Binary Black Holes}

\author[a,b]{Leonardo Iampieri}
\author[b]{Simone Mastrogiovanni}

\affiliation[a]{Dipartimento di Fisica, Sapienza Università di Roma,
I-00185 Roma, Italy}
\affiliation[b]{INFN, Sezione di Roma, I-00185 Roma, Italy}

\emailAdd{leonardo.iampieri@uniroma1.it}
\emailAdd{simone.mastrogiovanni@roma1.infn.it}

\abstract{
Low-dimensional parametric mass models are standard in gravitational-wave population
inference, but their rigidity can bias the recovered distribution and the conclusions
drawn from it.
We present a pipeline in which the source-frame binary-black-hole population in
$(z,m_1,m_2)$ is represented by a normalizing flow trained directly through the
hierarchical likelihood, with event posterior samples propagating measurement
uncertainty and detected injections accounting for selection effects.
We restrict population statements to the region supported by the injection campaign;
conditioning on this region can itself induce an apparent association between mass
and redshift.
We therefore introduce a mutual-information
diagnostic that compares the reconstruction with a redshift-independent reference under
the same support restriction, testing for dependence beyond that induced by the support
geometry.
We validate the method on two simulated catalogues differing only in whether the
characteristic primary-mass scale evolves with redshift.
The reconstruction recovers the injected mass structure in both, and the benchmarks
establish how the diagnostic behaves in the presence and absence of intrinsic
evolution.
Applied to the GWTC-5.0 catalogue, the method recovers features near $10\,M_\odot$
and $35\,M_\odot$, consistent with the LVK population analysis of the same catalogue,
and finds no evidence for intrinsic evolution of the primary-mass spectrum with
redshift.
More broadly, the pipeline addresses a problem common to many observational sciences:
recovering a population distribution from noisy, indirect, and selection-biased
measurements of its members.
}

\keywords{gravitational-wave population inference, binary black holes,
normalizing flows, hierarchical inference, selection effects}

\begin{document}

\maketitle
\flushbottom

\section{Introduction}
\label{sec:introduction}

The growing number of gravitational-wave (GW) detections has shifted compact-binary
astrophysics from the characterization of individual events to the inference of the
underlying population. In current LIGO--Virgo--KAGRA (LVK) analyses, the binary black
hole (BBH) source-frame mass distribution is typically described with low-dimensional
phenomenological families, such as truncated or broken power laws supplemented by
Gaussian-like peaks and a small number of parameters controlling slopes, cutoffs, and
peak properties \cite{LVKPopulationGWTC2,LVKPopulationGWTC3}. These models are
interpretable, computationally efficient, and straightforward to embed in hierarchical
Bayesian inference. Their limitation is that they commit to a restricted set of functional
forms before the data are examined. As catalogues grow, this rigidity can become a
source of systematic error if the true population departs from the assumed family,
especially when the data begin to resolve structure beyond the model's built-in
parametrization
\cite{Karathanasis2023BBHCosmology,Mastrogiovanni2021SourcePopCosmo,
Pierra2024Systematics,Rinaldi2024Evolution}.

This issue is relevant not only for astrophysical population studies, but also for GW
cosmology. Dark-siren methods use GW events without an identified electromagnetic
counterpart to constrain cosmological parameters, either statistically through galaxy
catalogues or through information carried by the population of detected compact binaries
\cite{SoaresSantos2019DarkSiren,Gray2020MockSirens,Gray2023JointPopCosmo}. In
spectral-siren analyses, the cosmology determines the mapping between detector-frame and
source-frame masses; consequently, assumptions about the intrinsic mass distribution can
provide redshift information and enter directly into the cosmological likelihood
\cite{Farr2019SpectralSirens,EzquiagaHolz2022SpectralSirens,Karathanasis2023BBHCosmology}.
In that setting, population misspecification becomes a source of systematic error:
an inflexible source-frame mass model can bias cosmological inference
\cite{Mastrogiovanni2021SourcePopCosmo,Pierra2024Systematics,Rinaldi2024Evolution}.
These considerations motivate population models that are flexible enough to capture
unanticipated structure without imposing it a priori.

In this work, we develop and validate a flexible data-driven BBH population model based on
normalizing flows (NFs) \cite{Rezende2015Flows,Papamakarios2019Flows,Kobyzev_2021}. A
normalizing flow represents a probability density through an invertible neural
transformation that maps a simple base distribution to a complex target distribution while
retaining exact density evaluation and efficient sampling. This makes NFs attractive as
flexible, data-driven density models for population inference: unlike low-dimensional
phenomenological families, the functional form can adapt to the data rather than being
fixed in advance by a small set of pre-chosen parameters.

The statistical setting is nevertheless different from ordinary density estimation.
The objects available for training are not exact draws from the target population:
each event enters through a noisy posterior distribution, and the observed catalogue is
distorted by the selection function of the detector network. Learning a high-capacity
density model in the presence of both measurement uncertainty and selection bias is
therefore the core methodological challenge. Several complementary machine-learning
approaches have addressed different parts of this problem. Flow-based generative models
have been explored as flexible models for GW populations
\cite{Wong2020FlowPopInference}. Machine-learning emulators, including normalizing
flows, have also been used to interpolate population-synthesis calculations or synthetic
astrophysical catalogues for subsequent astrophysical or cosmological inference
\cite{WongGerosa2019MLPopulation,Colloms2025,Scarpa2026}, while
neural-posterior-estimation and related flow-based methods have been developed for other
population and cosmological applications
\cite{Leyde2024NPE,Stachurski2024GWFlows}.

Ruhe et al.~\cite{Ruhe2022} placed a normalizing flow within hierarchical Bayesian
analysis to reconstruct the observed GWTC-3 population without correcting for selection
effects, and reported that optimization became highly unstable when the selection function
was included. Subsequent work clarified that unbiased inference of the observable
compact-binary population still requires an explicit treatment of selection effects, albeit
in a form different from the standard intrinsic-population formulation
\cite{Toubiana2026}. In a complementary direction, Payne and Thrane
\cite{PayneThrane2023} showed that maximizing the population likelihood over arbitrary
distributions leads to a discrete weighted sum of delta functions; Guttman et al.\
\cite{Guttman2026} subsequently applied this maximum-population-likelihood framework to
GWTC-4. Machine-learning methods have also been developed to accelerate the treatment
of selection effects \cite{Gerosa2020SelectionNN,TalbotThrane2022Malmquist}.

Here we take a different route: the normalizing flow itself represents the intrinsic
population density, and its parameters are optimized directly through the hierarchical
population likelihood, with per-event posterior samples propagating measurement uncertainty
and an injection campaign providing the selection correction
\cite{Mandel2019Selection,Mastrogiovanni2023IcaroGW}. In this sense, the paper is not
only about BBH population inference and its relevance for GW cosmology; it also presents
a practical pipeline for training expressive density models from noisy, indirect, and
selection-biased observations. To improve robustness near the optimum, we further
stabilize the reconstruction through an ensemble built from near-optimal checkpoints along
the training trajectory.

We validate the method on two simulated BBH catalogues that differ only in whether the
characteristic primary-mass scale evolves with redshift. The comparison tests the ability
of the hierarchical NF to distinguish evolving and redshift-independent populations in the
presence of measurement uncertainty and selection effects. We then apply the validated
pipeline to GWTC-5.0.

The rest of the paper is organized as follows.
Section~\ref{sec:statistical-framework} introduces the statistical framework: the
normalizing-flow population model and the hierarchical likelihood used throughout.
Section~\ref{sec:pipeline-validation} presents the analysis pipeline, the two
benchmark populations, and the validation on simulated catalogues.
Section~\ref{sec:gwtc5} applies the method to GWTC-5.0 and compares the recovered
mass structure and mass--redshift dependence with the literature.
Section~\ref{sec:conclusions} summarizes the findings and outlines extensions.
Technical details of the mock catalogues, the supported region, the training
configuration, and the Monte Carlo diagnostics are collected in the appendices.
\section{Statistical framework}
\label{sec:statistical-framework}

Population inference requires two complementary ingredients: a flexible representation of the latent population and a statistical procedure for learning that representation from imperfect observations. We use a normalizing flow to define a normalized and expressive density for the latent parameters, and a hierarchical likelihood to infer the flow parameters from event-level uncertainty while correcting for parameter-dependent selection effects. The construction is general and does not depend on a particular application; in the binary-black-hole analysis developed later, the latent variables are the source-frame parameters $\thv=(z,m_1,m_2)$.

The first subsection introduces the normalizing-flow population model at a general level, while the second explains how the flow density is inferred from posterior samples and injection campaigns. Technical details of the autoregressive neural-spline architecture, coordinate transformations, optimization settings, and Monte Carlo diagnostics are deferred to the appendix.

\subsection{Normalizing-flow population model}
\label{sec:normalizing-flow}

A normalizing flow is a flexible parametric family of normalized probability densities constructed by transforming a simple reference distribution through a sequence of invertible and differentiable maps \cite{Rezende2015Flows,Papamakarios2019Flows}.
We denote the resulting population density by $\qphi(\thv)$, where $\thv\in\mathbb{R}^{D}$ represents the latent parameters of an object and $\phi$ collects the trainable parameters of the transformations.
The combination of tractable density evaluation and efficient sampling makes normalizing flows particularly suitable for the present problem: the density must be evaluated at event-level posterior samples and simulated injections during inference, while samples from the reconstructed population are required for the subsequent population summaries.

Let $\bm u_0\in\mathbb{R}^{D}$ be distributed according to a tractable base density $p_0(\bm u_0)$.
A normalizing flow is defined by composing $K$ bijections,
\begin{equation}
  \bm u_k = f_k(\bm u_{k-1};\phi_k),
  \qquad
  k=1,\ldots,K,
  \qquad
  \thv \equiv \bm u_K,
  \label{eq:flow_comp_def}
\end{equation}
where each transformation $f_k$ is invertible and has a tractable Jacobian determinant.
Writing the full transformation as $f_{\phi}=f_K\circ\cdots\circ f_1$, the density induced on the target space follows from the change-of-variables formula,
\begin{align}
  \qphi(\thv)
  &=
  p_0\!\left(f_{\phi}^{-1}(\thv)\right)
  \left|
    \det
    \frac{\partial f_{\phi}^{-1}(\thv)}
         {\partial \thv}
  \right|
  \notag\\
  &=
  p_0\!\left(\bm u_0(\thv)\right)
  \prod_{k=1}^{K}
  \left|
    \det
    \frac{\partial f_k(\bm u_{k-1};\phi_k)}
         {\partial \bm u_{k-1}}
  \right|^{-1}.
  \label{eq:flow_changevar}
\end{align}
The first line evaluates the density by mapping a target-space point back to the base distribution, while the Jacobian factor accounts for the local change in volume produced by the transformation.
Sampling proceeds in the opposite direction by drawing $\bm u_0\sim p_0$ and applying the forward maps in Eq.~\eqref{eq:flow_comp_def}.

The expressiveness and computational cost of a flow are determined by the choice of invertible transformations.
In this work, we use an autoregressive neural spline flow \cite{papamakarios2017maf,Durkan2019NeuralSpline}.
The autoregressive factorization, rational-quadratic spline transformations, and architecture used in the analysis are described in Appendix~\ref{app:flow-details}; only the general properties of normalization, invertibility, and tractable density evaluation are required for the statistical development in the main text.

A normalizing flow alone does not specify how the population density should be learned from observations.
Ordinary density estimation would treat the available data as exact and unbiased draws from $\qphi$, whereas the catalogues considered here contain uncertain measurements and are distorted by parameter-dependent selection.
The following subsection therefore embeds the flow density in a hierarchical likelihood that propagates event-level uncertainty and corrects for selection effects.
\subsection{Hierarchical Bayesian inference}
\label{sec:hierarchical-inference}

The previous subsection introduced the normalizing-flow density $\qphi(\thv)$ used to represent the latent population.
We now describe how its parameters are inferred from a catalogue of noisy and selection-biased observations.
Let $\{\bm d_i\}_{i=1}^{C}$ denote the data associated with $C$ detected objects, and let $\thv$ denote their latent parameters.
In the binary-black-hole application considered later, $\bm d_i$ denotes the data associated with the $i$th event and $\thv=(z,m_1,m_2)$, but the derivation below does not depend on this particular choice of observables or latent parameters.

Conditioning on the observed catalogue size and focusing on the shape of the population, the hierarchical likelihood is, up to factors independent of $\phi$ \cite{Mandel2019Selection},
\begin{equation}
  \mathcal{L}(\{\bm d_i\}_{i=1}^{C}\mid\phi)
  \propto
  \prod_{i=1}^{C}
  \frac{
    \displaystyle
    \int p(\bm d_i\mid\thv)\,\qphi(\thv)\,\dd\thv
  }{
    \alpha(\phi)
  },
  \label{eq:hierlike}
\end{equation}
where $p(\bm d_i\mid\thv)$ is the likelihood for a single object.
The selection factor is
\begin{equation}
  \alpha(\phi)
  \equiv
  \int \pdet(\thv)\,\qphi(\thv)\,\dd\thv,
  \label{eq:alpha_def}
\end{equation}
where $\pdet(\thv)$ is the probability that an object with parameters $\thv$ satisfies the detection and analysis criteria.
The event integrals in the numerator propagate measurement uncertainty, while the factor $\alpha(\phi)^{-C}$ corrects for the parameter-dependent selection of the observed catalogue.
An analysis that also inferred the absolute occurrence rate would include the corresponding count term.

\paragraph{Event-level uncertainty.}

The event integrals in Eq.~\eqref{eq:hierlike} can be evaluated using posterior samples without repeatedly evaluating the original data likelihood.
Suppose that the analysis of object $i$ provides samples $\{\thv_{i,k}\}_{k=1}^{K_i}$ from the posterior
\begin{equation}
  p(\thv\mid\bm d_i)
  =
  \frac{
    p(\bm d_i\mid\thv)\,\pi_i(\thv)
  }{
    p(\bm d_i)
  },
  \label{eq:event_bayes}
\end{equation}
obtained under an interim prior $\pi_i(\thv)$.
Rewriting the event integral as an expectation over this posterior gives
\begin{align}
  \int p(\bm d_i\mid\thv)\,\qphi(\thv)\,\dd\thv
  &\propto
  \int p(\thv\mid\bm d_i)
  \frac{\qphi(\thv)}{\pi_i(\thv)}
  \,\dd\thv
  \notag\\
  &\approx
  \frac{1}{K_i}
  \sum_{k=1}^{K_i}
  \frac{
    \qphi(\thv_{i,k})
  }{
    \pi_i(\thv_{i,k})
  }.
  \label{eq:event_mc}
\end{align}
The evidence $p(\bm d_i)$ is omitted because it does not depend on the population model.
Each object therefore contributes through an average of population-to-interim-prior importance weights over its posterior samples.
This construction propagates broad and non-Gaussian event-level uncertainty without reducing each observation to a single point estimate.

\paragraph{Selection effects.}

The selection factor is evaluated directly from a Monte Carlo injection campaign.
Let $N_{\rm tot}^{\rm gen}$ parameter values be generated from an injection density $\pinj(\thv)$, and let $\delta_j\in\{0,1\}$ indicate whether injection $j$ passes the same detection and analysis criteria used to construct the catalogue.
Then
\begin{equation}
  \alpha(\phi)
  \approx
  \frac{1}{N_{\rm tot}^{\rm gen}}
  \sum_{j=1}^{N_{\rm tot}^{\rm gen}}
  \delta_j
  \frac{
    \qphi(\thv_j^{\rm inj})
  }{
    \pinj(\thv_j^{\rm inj})
  }
  =
  \frac{1}{N_{\rm tot}^{\rm gen}}
  \sum_{j=1}^{N_{\rm det}^{\rm inj}}
  \frac{
    \qphi(\thv_j^{\rm inj})
  }{
    \pinj(\thv_j^{\rm inj})
  },
  \label{eq:alpha_mc}
\end{equation}
where the final sum runs only over detected injections \cite{Mandel2019Selection,Mastrogiovanni2023IcaroGW}.
Equation~\eqref{eq:alpha_mc} uses the injection campaign directly and does not require a separately fitted model of $\pdet(\thv)$.

Combining Eqs.~\eqref{eq:event_mc} and \eqref{eq:alpha_mc}, the Monte Carlo approximation to the shape-only log-likelihood, up to terms independent of $\phi$, is
\begin{align}
  \log \widehat{\mathcal L}(\phi)
  ={}&
  \sum_{i=1}^{C}
  \log\!\left[
    \frac{1}{K_i}
    \sum_{k=1}^{K_i}
    \frac{
      \qphi(\thv_{i,k})
    }{
      \pi_i(\thv_{i,k})
    }
  \right]
  \notag\\
  &-
  C\log\!\left[
    \frac{1}{N_{\rm tot}^{\rm gen}}
    \sum_{j=1}^{N_{\rm det}^{\rm inj}}
    \frac{
      \qphi(\thv_j^{\rm inj})
    }{
      \pinj(\thv_j^{\rm inj})
    }
  \right].
  \label{eq:mc_hier_loglike}
\end{align}
This expression is the objective from which the flow density is learned.
If posterior samples and injections are supplied in coordinates different from those used by the flow, the interim and injection prior densities must be transformed with the corresponding Jacobians so that every ratio in Eq.~\eqref{eq:mc_hier_loglike} is defined with respect to the same measure.

We assess the numerical reliability of the posterior-sample and injection-based Monte Carlo estimates using effective-sample-size diagnostics.
Their definitions and the values obtained in our analyses are reported in Appendix~\ref{app:monte-carlo-diagnostics}.

The framework applies whenever latent populations are observed indirectly, individual measurements are uncertain, and inclusion in the catalogue depends on the latent parameters.
In this work, we optimize the hierarchical likelihood with respect to the flow parameters $\phi$ rather than sampling a full Bayesian posterior over the neural-network weights.
The ensemble introduced in Section~\ref{sec:pipeline-validation} is therefore used to assess reconstruction robustness and should not be interpreted as a Bayesian credible region for the population density.

\FloatBarrier
\section{Pipeline presentation and validation}
\label{sec:pipeline-validation}

Having defined the normalizing-flow population model and the hierarchical
likelihood, we now combine them into the analysis pipeline used throughout this
paper and validate it on controlled simulations.
We first describe the pipeline in general terms: the latent population is
reconstructed with a cross-validated ensemble of hierarchically trained
normalizing flows, the ensemble density is interpreted within the
injection-supported region $\Stilde$, and its mass--redshift dependence is
quantified with the mutual-information diagnostic introduced below.
We then introduce two benchmark populations that differ only in whether the
primary-mass distribution evolves with redshift, and apply the complete procedure
to the corresponding simulated catalogues, comparing the recovered distributions
and the diagnostic values with the known injected populations.
Only the general benchmark design is retained in the main text; the detailed
construction of the mock detections, event-level posterior samples, and injection
campaigns is given in Appendix~\ref{app:mock-catalogues}.

\begin{figure}[t]
  \centering
  \includegraphics[
    width=\linewidth
  ]{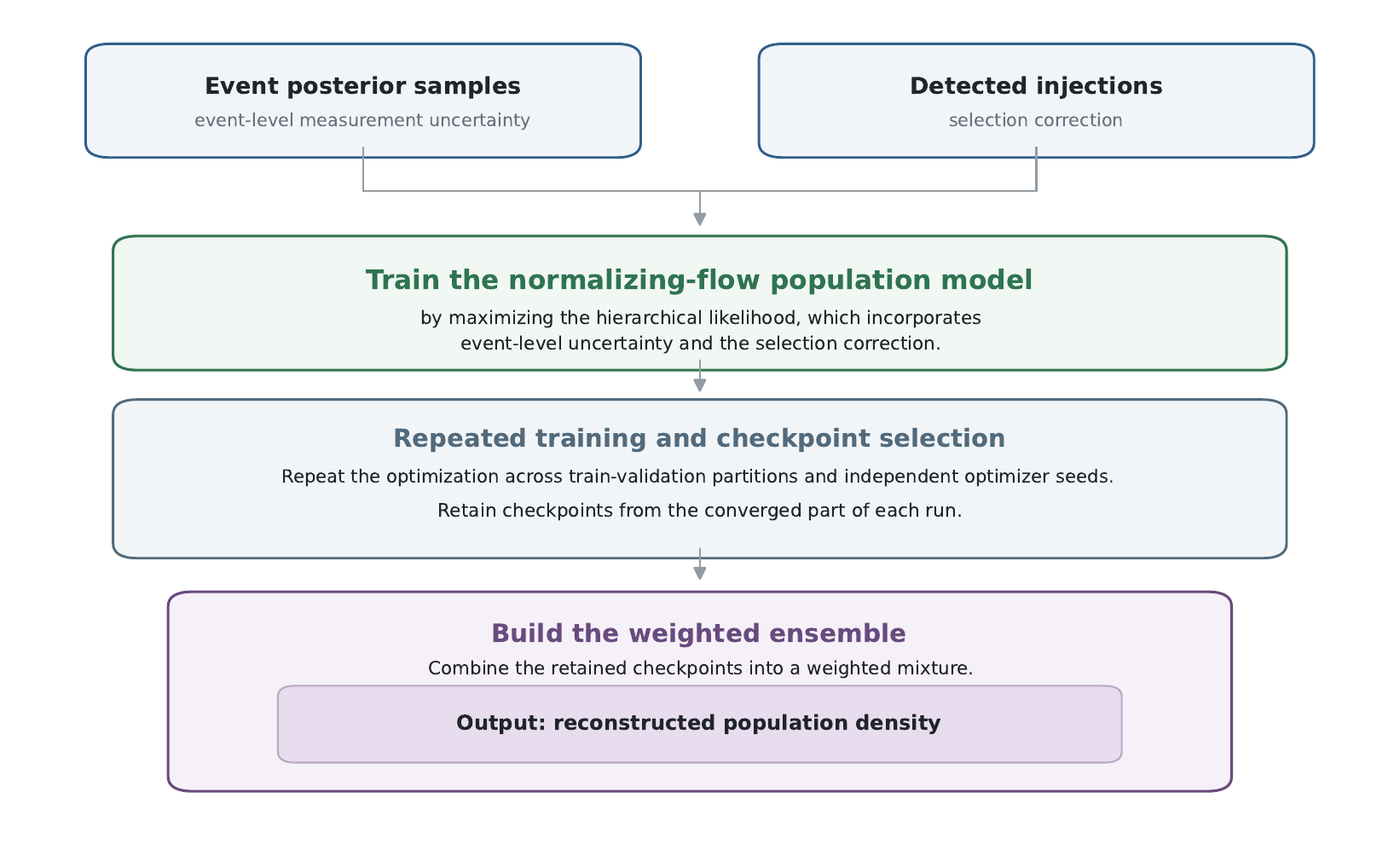}
  \caption{%
  Schematic overview of the population-reconstruction pipeline.
  Event posterior samples propagate measurement uncertainty, while detected
  injections provide the selection correction.
  The normalizing-flow population model is trained through the hierarchical
  likelihood across repeated train--validation runs, from which checkpoints in
  the converged part of each trajectory are retained.
  The retained checkpoints are combined into a weighted ensemble that defines
  the reconstructed population density.
  }
  \label{fig:pipeline-overview}
\end{figure}

Starting from event-level posterior samples and a detected-injection campaign, we
reconstruct the latent population with a cross-validated ensemble of normalizing
flows trained through the hierarchical likelihood.
The retained checkpoints from the training runs are combined into a weighted
ensemble, whose mixture defines our central population-density estimate.
The reconstruction procedure is summarized in Fig.~\ref{fig:pipeline-overview}.

We report the inferred population only within the region supported by the detected
injections, and assess mass--redshift dependence by comparing the reconstruction
with a redshift-independent reference under the same support restriction.
The same reconstruction, support restriction, and diagnostic procedure is applied
to the benchmark simulations and to the observational catalogue.
The corresponding joint and redshift-resolved population summaries are presented
together with those results.

\paragraph{Population reconstruction.}
To improve numerical conditioning, the population parameters $\thv$ are mapped to
standardized logarithmic coordinates
\begin{equation}
  \bm u
  =
  s\,\frac{\log\thv-\bm\mu}{\bm\sigma},
  \label{eq:u_def_training}
\end{equation}
where the logarithm and division act component-wise, $\bm\mu$ and $\bm\sigma$ are
the component-wise mean and standard deviation of $\log\thv$ over the catalogue
posterior samples, and the scale $s$ places the transformed samples within the
stable domain of the spline transformations.
The same map is applied to posterior samples and injections, and the interim and
injection prior densities are transformed through it, including its Jacobian, so
that every importance ratio is evaluated with respect to a common measure.
When the supplied samples are given in different variables, as for the
detector-frame binary-black-hole data used here, the map also includes the
transformation to $\thv$ and the corresponding Jacobian.

For a subset of events $\mathcal D$ containing $N_{\mathcal D}$ objects, the flow
parameters $\phi$ are obtained by minimizing the mean per-event hierarchical
negative log-likelihood
\begin{equation}
  \bar{\mathcal L}_{\mathcal D}(\phi)
  =
  -\frac{1}{N_{\mathcal D}}
  \sum_{i\in\mathcal D}
  \log\!\left[
    \frac{1}{K_i}\sum_{k=1}^{K_i}
    \frac{\qphi(\bm u_{i,k})}{\pi_{i,u}(\bm u_{i,k})}
  \right]
  +
  \log\!\left[
    \frac{1}{N_{\rm tot}^{\rm gen}}
    \sum_{j=1}^{N_{\rm det}^{\rm inj}}
    \frac{\qphi(\bm u_j^{\rm inj})}{\pi_{{\rm inj},u}(\bm u_j^{\rm inj})}
  \right],
  \label{eq:mean_hier_loss}
\end{equation}
which is the negative Monte Carlo hierarchical log-likelihood of
Eq.~\eqref{eq:mc_hier_loglike}, divided by the number of events: the first term
propagates event-level uncertainty through posterior-sample reweighting, and the
second applies the injection-derived selection correction.
Dividing by $N_{\mathcal D}$ leaves the optimum unchanged for a fixed catalogue and
keeps the loss on a comparable scale across training and validation evaluations.

The flow is trained with Adam on mini-batches of catalogue events.
For each optimizer step, $\mathcal D$ in Eq.~\eqref{eq:mean_hier_loss} is a random
mini-batch $\mathcal B$ of $B$ events drawn from the training set.
For each event in the mini-batch the event integral is estimated from a random
subset of its posterior samples, and the selection factor from a random subset of
the detected injections, so that successive updates see independent Monte Carlo
realizations of both terms in Eq.~\eqref{eq:mean_hier_loss}.

Training is organized through four independent five-fold partitions of the
catalogue.
Within each partition, every fold is used once for validation while the remaining
events are used for optimization.
The validation loss $\bar{\mathcal L}_{\rm val}$, obtained by evaluating
Eq.~\eqref{eq:mean_hier_loss} on the validation fold, controls learning-rate
adaptation and stopping.
The flow architecture, batch sizes, optimizer schedule, and remaining training
details are given in Appendix~\ref{app:training-details}. The numerical tests
reported there show that, at the adopted settings, updates computed from finite
event batches remain closely aligned with the full-catalogue update. They also
show that posterior-sample and injection subsampling contribute less to the
variability of an update than the composition of the event batch.

\paragraph{Ensemble reconstruction.}
We retain the converged part of each optimization trajectory rather than a single
final checkpoint.
Every saved checkpoint is evaluated on its run's validation fold using fixed
high-accuracy Monte Carlo settings.
Writing $\bar{\mathcal L}^{(r,t)}_{\rm val}$ for the validation loss of checkpoint
$t$ in run $r$, we retain the checkpoints satisfying
\begin{equation}
  \bar{\mathcal L}^{(r,t)}_{\rm val}
  -
  \min_{t'}\bar{\mathcal L}^{(r,t')}_{\rm val}
  \le
  \Delta_{\rm cut},
  \label{eq:delta_plateau}
\end{equation}
where $\Delta_{\rm cut}$ is chosen separately for each analysis to keep
reconstructions from the near-optimal part of the trajectory.

With $\mathcal A_r$ the checkpoints retained from run $r$ and $R_{\rm eff}$ the
number of runs retaining at least one, the reconstruction is the weighted mixture
\begin{equation}
  q_{\rm ens}(\bm u)
  =
  \frac{1}{R_{\rm eff}}
  \sum_{r=1}^{R_{\rm eff}}
  \sum_{a\in\mathcal A_r}
  \omega_{a\mid r}\,q_{\phi_a}(\bm u),
  \qquad
  \omega_{a\mid r}
  =
  \frac{\exp[-\bar{\mathcal L}_{{\rm val},a}]}
       {\sum_{b\in\mathcal A_r}
        \exp[-\bar{\mathcal L}_{{\rm val},b}]}.
  \label{eq:weighted_ensemble}
\end{equation}
Each run contributes equal total weight, so that a run cannot dominate merely by
retaining more checkpoints, while within a run the normalized exponential
weights favour checkpoints with lower validation loss.

The weighted mixture $q_{\rm ens}$ is the central population-density estimator.
It is shown in the figures together with the spread across retained members,
which quantifies the sensitivity of the reconstruction to the train--validation
partition, the optimizer seed, and residual variation along the converged
trajectory.
This spread is a robustness diagnostic rather than a Bayesian credible region. Mapping the ensemble back to the physical population coordinates gives
\begin{equation}
  p_{\rm rec}(\thv)
  =
  q_{\rm ens}\!\left(\bm u(\thv)\right)
  \left|\det\frac{\partial\bm u}{\partial\thv}\right|,
  \label{eq:ensemble_physical_density}
\end{equation}
which is the reconstructed density used, in physical coordinates, throughout the
remainder of the analysis.

\paragraph{Region of support.}
The reconstruction is trustworthy only where the detected injections support the
Monte Carlo selection correction; outside that region the flexible flow is
unconstrained.
This domain is the support of the detected-injection distribution,
\begin{equation}
  \mathcal S
  \equiv
  \supp\!\left(\pdet(\thv)\,\pinj(\thv)\right)
  =
  \left\{\thv:\pdet(\thv)\,\pinj(\thv)>0\right\},
  \label{eq:support_def}
\end{equation}
which, being available only through a finite injection sample, we approximate by
a conservative data-driven region $\Stilde$ obtained by intersecting several
split-calibrated nearest-neighbour regions built from the detected injections in
the standardized coordinates of Eq.~\eqref{eq:u_def_training} with the physical
parameter constraints; its construction is given in Appendix~\ref{app:support}.
The ensemble is built before this restriction is applied; all reported densities,
projections, and comparisons are then obtained by conditioning $p_{\rm rec}$ on the same
region $\Stilde$.

\paragraph{Mass--redshift dependence.}
The boundary of $\Stilde$ depends on both redshift and mass, and can induce an
apparent association between them even when the intrinsic population is
redshift independent.
To distinguish this geometric effect from a dependence inferred from the data,
we compare the reconstructed physical density $p_{\rm rec}(z,m_1,m_2)$ of
Eq.~\eqref{eq:ensemble_physical_density} with a redshift-independent reference
under the same support restriction.

The diagnostic is designed to probe evolution of the primary-mass spectrum.
The reconstruction can also contain redshift dependence in the secondary-mass
distribution, which may affect the $(z,m_1)$ projection once the
three-dimensional support is imposed.
We remove this contribution by replacing
$p_{\rm rec}(m_2\mid z,m_1)$ with the corresponding conditional
$p_{\rm rec}(m_2\mid m_1)$ before applying the support restriction.
The reconstructed $(z,m_1)$ distribution itself is therefore left unchanged.

After conditioning on $\Stilde$ and marginalizing over $m_2$, the reconstruction
and its redshift-independent reference are
\begin{align}
  P_{\rm rec}(z,m_1\mid\Stilde)
  &\propto
  p_{\rm rec}(z,m_1)
  \int
  \mathbf{1}_{\Stilde}(z,m_1,m_2)\,
  p_{\rm rec}(m_2\mid m_1)\,
  \dd m_2,
  \nonumber\\
  P_{\rm ref}(z,m_1\mid\Stilde)
  &\propto
  p_{\rm rec}(z)\,p_{\rm rec}(m_1)
  \int
  \mathbf{1}_{\Stilde}(z,m_1,m_2)\,
  p_{\rm rec}(m_2\mid m_1)\,
  \dd m_2,
  \label{eq:redshift_independent_reference}
\end{align}
with each density normalized over $(z,m_1)$.
All marginals and conditional distributions appearing here are obtained from
$p_{\rm rec}$ before conditioning on $\Stilde$.
The two densities therefore differ only in whether the reconstructed redshift
and primary mass remain coupled; the secondary-mass distribution and support
restriction are the same in both.
We restrict the primary mass to
\begin{equation}
  25\,M_\odot
  \le
  m_1
  <
  80\,M_\odot,
  \label{eq:deltaI-m1-range}
\end{equation}
which encompasses the higher-mass feature reported in the GWTC-5.0 population
analysis \cite{GWTC5Population}.

For either density, the dependence between redshift and primary mass is measured
by the mutual information
\begin{equation}
  I_x
  =
  \int
  P_x(z,m_1\mid\Stilde)\,
  \ln\!\left[
    \frac{P_x(z,m_1\mid\Stilde)}
         {P_x(z\mid\Stilde)\,P_x(m_1\mid\Stilde)}
  \right]
  \dd z\,\dd m_1,
  \qquad
  x\in\{{\rm rec},{\rm ref}\},
  \label{eq:supported_mutual_information}
\end{equation}
and we define the mass--redshift diagnostic as the signed difference
\begin{equation}
  \DeltaIS
  =
  I_{\rm rec}-I_{\rm ref}.
  \label{eq:global_evolution_diagnostic}
\end{equation}
The diagonal $I_{\rm rec}=I_{\rm ref}$ corresponds to $\DeltaIS=0$.
Retaining the sign allows fluctuations around the redshift-independent reference
to appear on either side of zero rather than being folded onto positive values.
If the reconstructed $(z,m_1)$ distribution factorizes exactly, the two
supported densities coincide and the diagnostic vanishes.
This is a diagnostic of the reconstructed density rather than a formal Bayesian
comparison between evolving and non-evolving populations.
It is first evaluated on the two benchmark populations and then applied without
modification to GWTC-5.0.
\subsection{Benchmark populations}
\label{sec:benchmark-populations}

We validate the pipeline on two controlled binary-black-hole populations that share the same redshift distribution, mass spectrum, detection model, and mock-measurement procedure.
The only difference between them is whether the characteristic peak in the primary-mass distribution remains fixed or shifts with redshift.
This construction isolates the effect that the pipeline is intended to recover: intrinsic mass--redshift dependence in the presence of measurement uncertainty and selection effects.

Each source is described by the source-frame parameters
\begin{equation}
  \thv=(z,m_1,m_2),
  \qquad
  m_1\geq m_2,
  \label{eq:benchmark_parameters}
\end{equation}
and is drawn from the factorized population density
\begin{equation}
  p_{\rm pop}(z,m_1,m_2)
  =
  p(z)\,
  p(m_1\mid z)\,
  p(m_2\mid m_1).
  \label{eq:benchmark_factorization}
\end{equation}

The redshift distribution is
\begin{equation}
  p(z)
  \propto
  \frac{\dd V_c}{\dd z}\,
  \frac{R(z)}{1+z},
  \label{eq:benchmark_redshift_distribution}
\end{equation}
where $\dd V_c/\dd z$ is the differential comoving volume element and the factor $(1+z)^{-1}$ accounts for cosmological time dilation.
We evaluate the comoving volume in a flat $\Lambda$CDM cosmology with
\begin{equation}
  H_0=67.7~{\rm km\,s^{-1}\,Mpc^{-1}},
  \qquad
  \Omega_m=0.3065,
\end{equation}
and use the Madau--Dickinson merger-rate evolution \cite{MadauDickinson2014},
\begin{equation}
  R(z)
  \propto
  \left[
    1+(1+z_p)^{-\gamma-\kappa}
  \right]
  \frac{
    (1+z)^\gamma
  }{
    1+
    \left(
      \frac{1+z}{1+z_p}
    \right)^{\gamma+\kappa}
  },
  \label{eq:benchmark_rate}
\end{equation}
with $(\gamma,\kappa,z_p)=(3,3,2)$.
The distribution is truncated to $z\in[10^{-3},10]$; the upper limit is only a numerical cutoff, since the probability of detecting sources at such large redshift is negligible for the mock selection model used here.

The primary-mass distribution is a mixture of a power law and a Gaussian component,
\begin{equation}
  p(m_1\mid z)
  =
  (1-\lambda_{\rm peak})\,
  p_{\rm PL}(m_1)
  +
  \lambda_{\rm peak}\,
  p_{\rm G}(m_1\mid z),
  \label{eq:benchmark_primary_mass}
\end{equation}
with
\begin{align}
  p_{\rm PL}(m_1)
  &\propto
  m_1^{-\alpha}\,
  \mathbb{I}(m_{\min}\leq m_1\leq m_{\max}),
  \label{eq:benchmark_power_law}
  \\
  p_{\rm G}(m_1\mid z)
  &\propto
  \exp\!\left[
    -\frac{
      \left(m_1-\mu_g(z)\right)^2
    }{
      2\sigma_g^2
    }
  \right]
  \mathbb{I}(m_{\min}\leq m_1\leq m_{\max}).
  \label{eq:benchmark_gaussian}
\end{align}
Both components are multiplied by the standard low-mass smoothing function used in binary-black-hole population analyses \cite{LVKPopulationGWTC3}.
We fix
\begin{equation}
  \alpha=3.78,
  \qquad
  m_{\min}=4.98\,M_\odot,
  \qquad
  m_{\max}=112.5\,M_\odot,
\end{equation}
and
\begin{equation}
  \delta_m=4.8\,M_\odot,
  \qquad
  \sigma_g=3.88\,M_\odot,
  \qquad
  \lambda_{\rm peak}=0.03.
\end{equation}

The Gaussian-peak mean is allowed to vary linearly with redshift,
\begin{equation}
  \mu_g(z)
  =
  \mu_{g,0}
  +
  \mu_{g,1}z.
  \label{eq:benchmark_peak_evolution}
\end{equation}
The two benchmark populations are then defined by
\begin{align}
  \text{Model A:}\qquad
  &\mu_{g,0}=32.27\,M_\odot,
  &
  \mu_{g,1}=0,
  \label{eq:model_a_definition}
  \\
  \text{Model B:}\qquad
  &\mu_{g,0}=32.27\,M_\odot,
  &
  \mu_{g,1}=5\,M_\odot.
  \label{eq:model_b_definition}
\end{align}
Model~A therefore has a redshift-independent primary-mass spectrum, whereas in Model~B the Gaussian feature moves to larger masses with increasing redshift.
All remaining population ingredients are identical, so differences between the two validation cases can be attributed directly to the injected peak evolution.

For the secondary mass, we adopt a conditional power law in the mass ratio $q=m_2/m_1$,
\begin{equation}
  p(m_2\mid m_1)
  \propto
  q^\beta\,
  \mathbb{I}(m_{\min}\leq m_2\leq m_1),
  \qquad
  q=\frac{m_2}{m_1},
  \label{eq:benchmark_secondary_mass}
\end{equation}
with $\beta=0.81$ and the same low-mass smoothing scale $\delta_m$ used for the primary mass.

\IfFileExists{figures/corner_true_ModelA_vs_ModelB_physical_axes_KDE.pdf}{%
\begin{figure}[t]
  \centering
  \includegraphics[
    width=0.9\linewidth
  ]{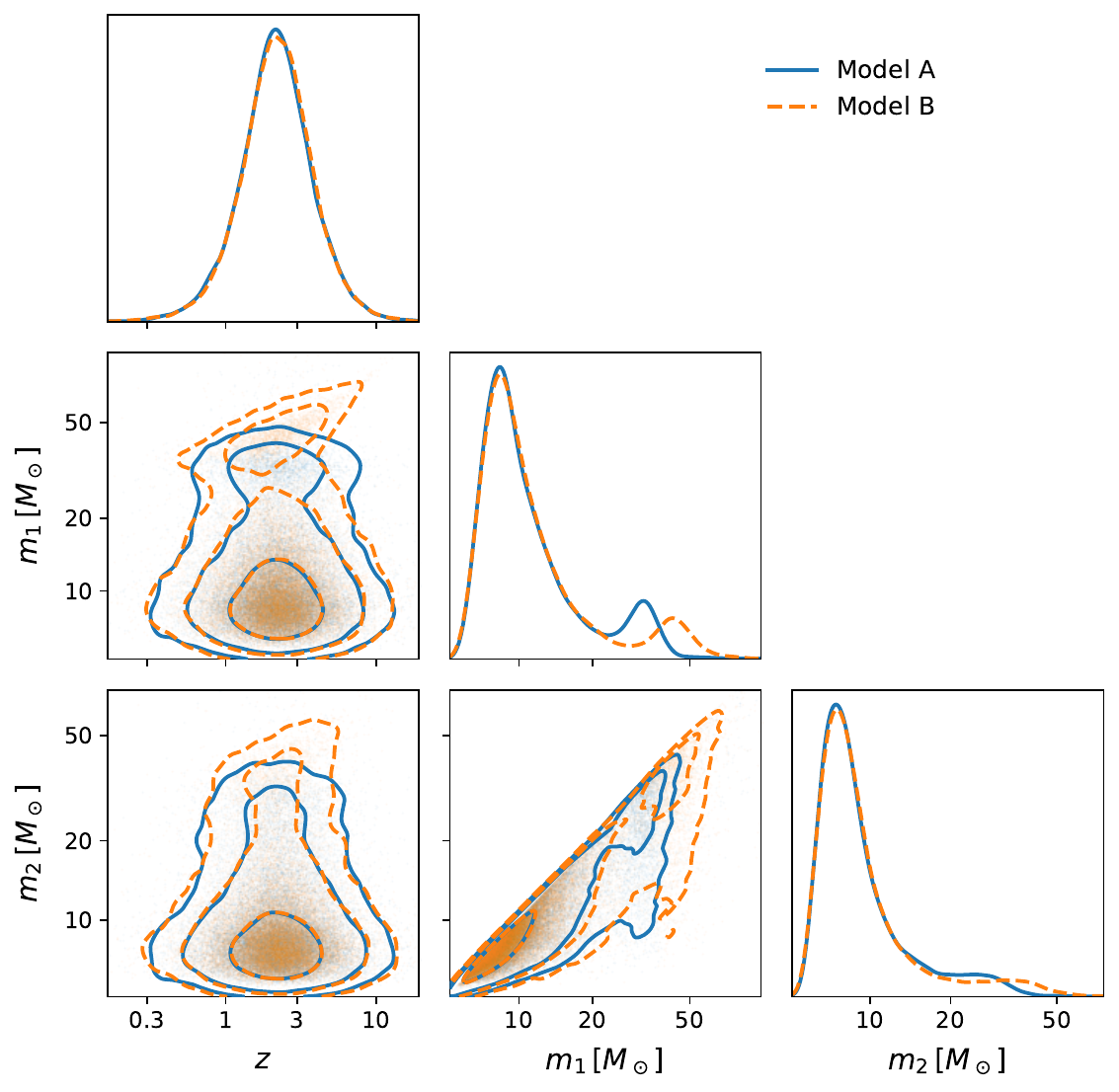}
  \caption{%
  Intrinsic source-frame distributions of the two benchmark populations.
  Model~A has a redshift-independent Gaussian feature in the primary-mass distribution, while in Model~B the mean of the feature increases linearly with redshift according to Eq.~\eqref{eq:benchmark_peak_evolution}.
  The redshift distribution, secondary-mass model, and all remaining population parameters are identical.
  }
  \label{fig:bench_corner}
\end{figure}
}{}

For each population, we generate a mock detected catalogue, synthetic event-level posterior samples, and a detected-injection campaign.
The same cosmology, measurement model, and detection rule are used for Models~A and~B.
The likelihood approximation, signal-to-noise-ratio selection, posterior-sample construction, implied sampling priors, and injection distribution are described in Appendix~\ref{app:mock-catalogues}.
The following subsection applies the complete reconstruction pipeline to these two catalogues and compares the inferred populations with the known injected distributions.
\subsection{Validation on simulated catalogues}
\label{sec:simulation-validation}

We now apply the complete pipeline to the two benchmark catalogues introduced in
Section~\ref{sec:benchmark-populations}.
The flows are trained and evaluated in the source-frame coordinates
$\thv=(z,m_1,m_2)$, in the same flat-$\Lambda$CDM cosmology used to generate the
simulations.
The construction of the mock detections, event-level posterior samples, interim
priors, and injection campaign is described in Appendix~\ref{app:mock-catalogues};
the flow architecture, batch sizes, optimization schedule, and checkpoint settings
are given in Appendix~\ref{app:training-details}.

\begin{figure}[t]
  \centering
  \begin{minipage}[t]{0.78\linewidth}
    \centering
    \includegraphics[
      width=\linewidth
    ]{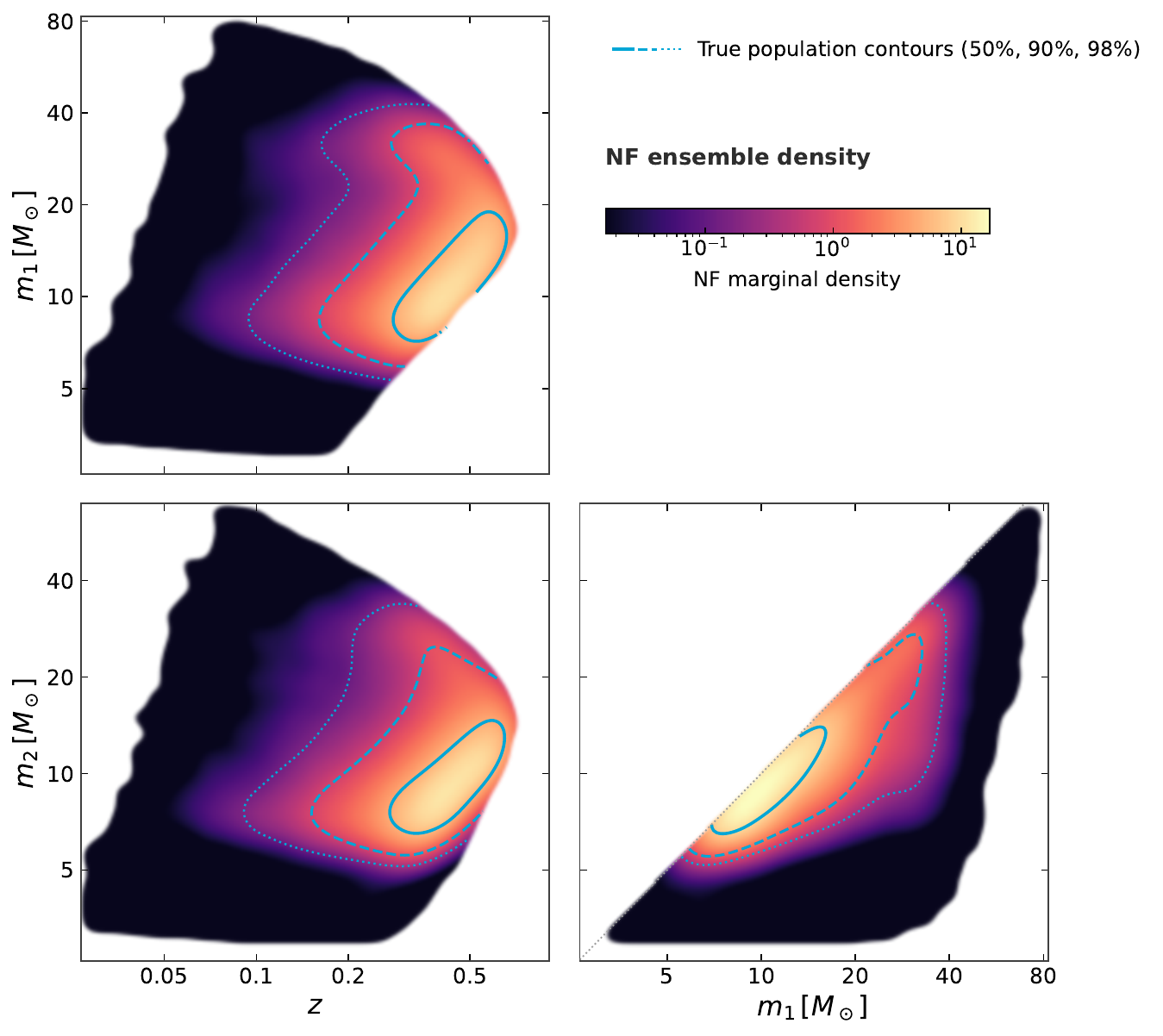}
    \smallskip
    (a) Model~A
  \end{minipage}

  \medskip

  \begin{minipage}[t]{0.78\linewidth}
    \centering
    \includegraphics[
      width=\linewidth
    ]{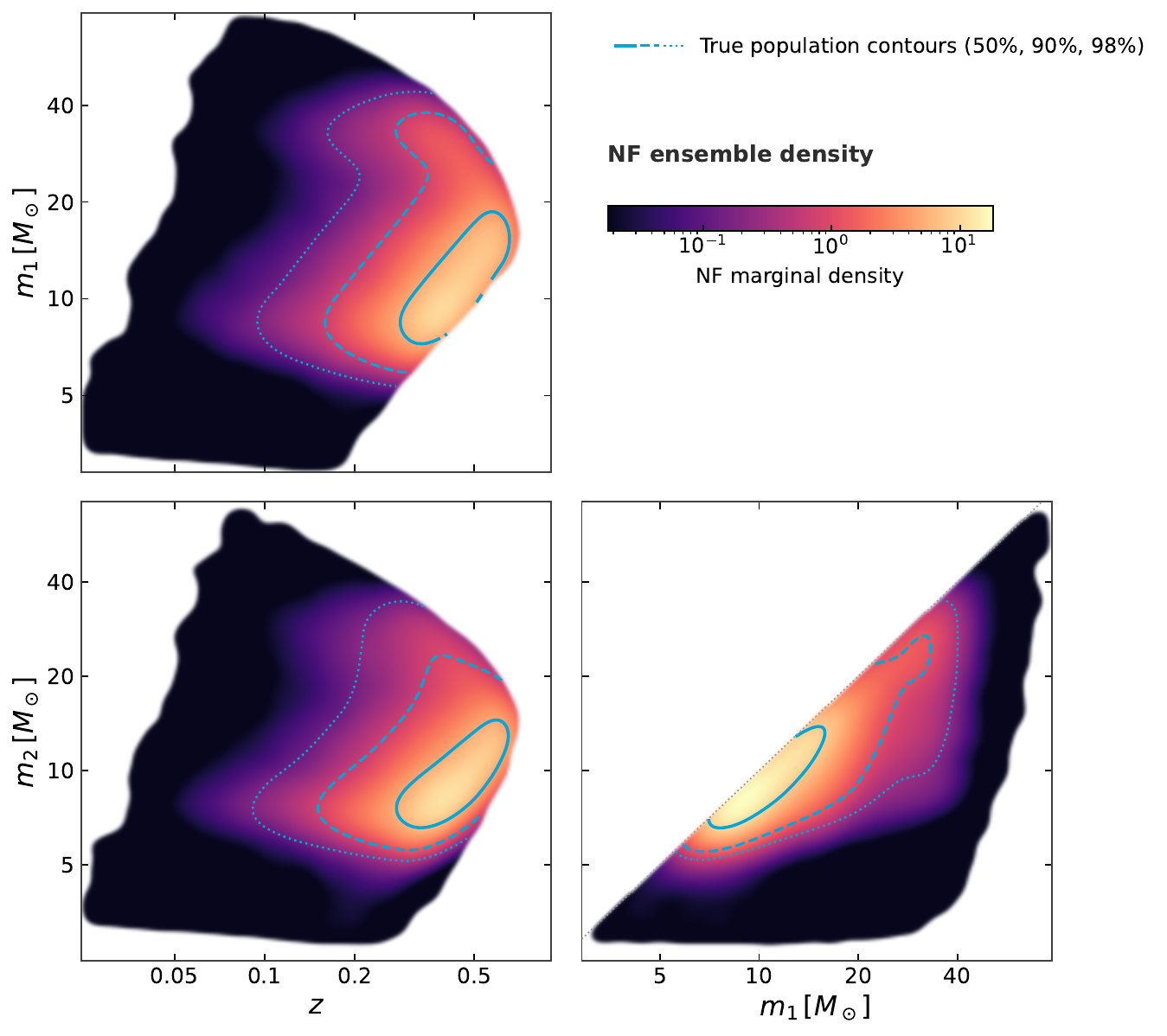}
    \smallskip
    (b) Model~B
  \end{minipage}

  \caption{%
  Reconstruction of the two benchmark populations within the injection-supported region of the $(z,m_1,m_2)$ parameter space.
  The colour scale shows the weighted normalizing-flow ensemble density, while
  the overlaid contours show the corresponding true population at the
  $50\%$, $90\%$, and $98\%$ enclosed probability levels.
  }
  \label{fig:benchmark-joint-reconstruction}
\end{figure}

As an overall view of the reconstruction,
Fig.~\ref{fig:benchmark-joint-reconstruction} compares the inferred population
with the corresponding true distribution across all three variables for
Models~A and B.
The ensemble reproduces the main structure and characteristic features of the
injected populations within the supported region.

\paragraph{Reconstruction in redshift bins.}
Figure~\ref{fig:benchmark-z-bins} shows the conditional primary-mass distributions
$p(\log_{10}(m_1/M_\odot)\mid z\in Z_b,\Stilde)$ in a fixed set of redshift
intervals $Z_b$, identical for the two populations.
In each bin the solid curve is the weighted ensemble mixture of
Eq.~\eqref{eq:weighted_ensemble} and the shaded band the central $90\%$ spread
across the retained members.
For both models the reconstruction recovers the injected mass features at the
correct primary masses, and the true conditional density generally lies within the ensemble spread. Residual differences between the ensemble and the true distributions in both models are consistent with the finite simulated catalogue, the finite posterior and
injection Monte Carlo samples, and the maximum-likelihood reconstruction within a
finite flow family.

The conditional distributions vary with redshift in both models, including Model~A,
whose source population does not evolve.
The binned reconstructions alone therefore cannot distinguish an intrinsic
mass--redshift dependence from one induced by the redshift-dependent geometry of
$\Stilde$.

\IfFileExists{figures/m1_z_slices_F1_FINAL_PAPER_ALIGNED.pdf}{%
\IfFileExists{figures/m1_z_slices_B_7_FINAL_PAPER_ALIGNED.pdf}{%
\begin{figure}[t]
  \centering
  \begin{minipage}[t]{0.49\linewidth}
    \centering
    \includegraphics[width=\linewidth]{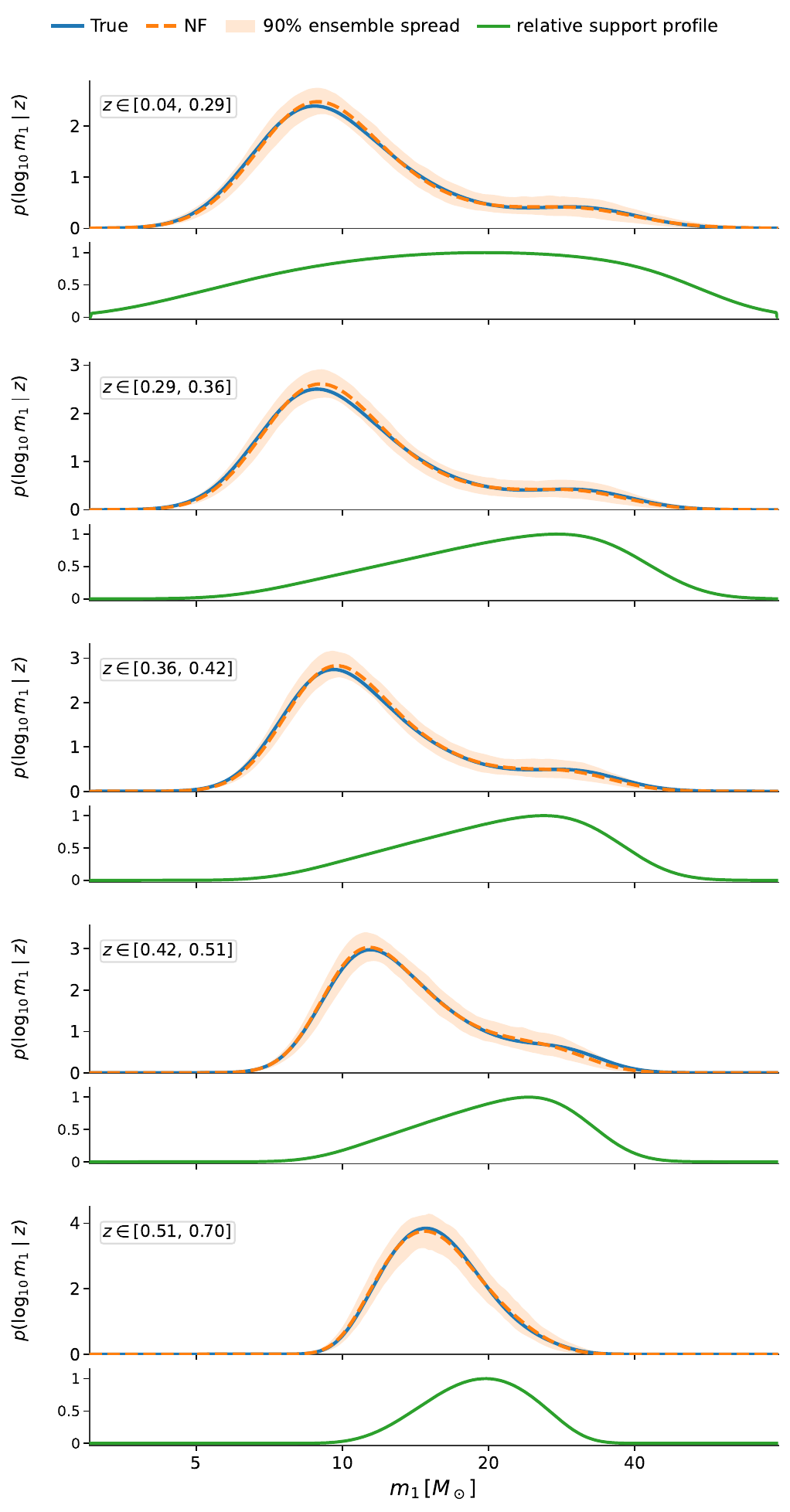}
    \smallskip
    (a) Model~A
  \end{minipage}\hfill
  \begin{minipage}[t]{0.49\linewidth}
    \centering
    \includegraphics[width=\linewidth]{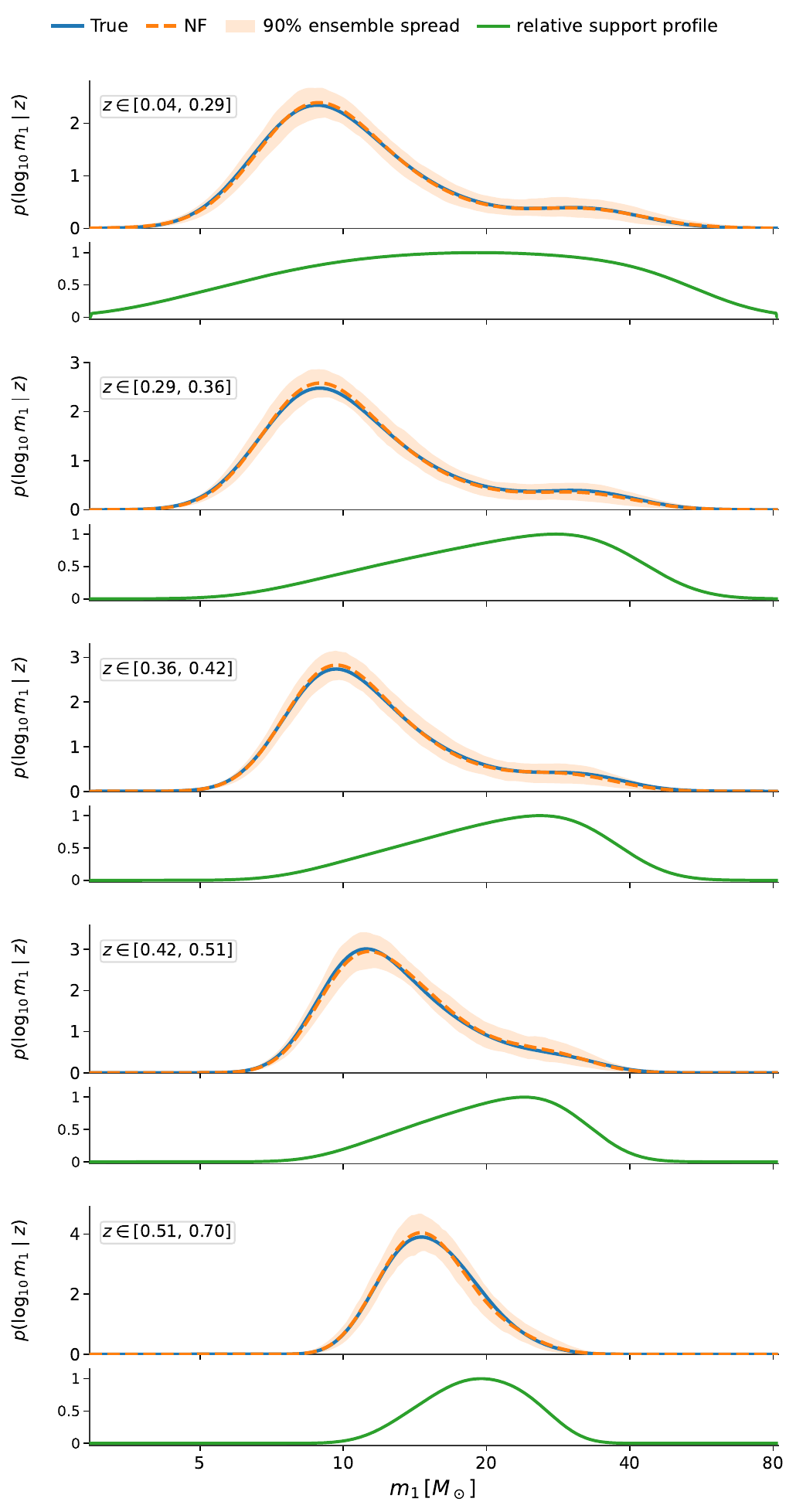}
    \smallskip
    (b) Model~B
  \end{minipage}
  \caption{%
  Redshift-resolved primary-mass reconstructions for the two benchmark
  populations, conditioned on the injection-supported region $\Stilde$.
  In each redshift bin the true conditional density in
  $\log_{10}(m_1/M_\odot)$ is compared with the weighted ensemble mixture and the central $90\%$ spread across retained members.
  The lower subpanels show the extent of the supported region as a function of
  primary mass---the fraction of the $(z,m_2)$ volume within the bin that lies
  inside $\Stilde$ at each $m_1$, normalized to its maximum in that bin---and
  therefore indicate where in $m_1$ the reconstruction is constrained.
  The construction of $\Stilde$ is described in
  Appendix~\ref{app:support}.
  }
  \label{fig:benchmark-z-bins}
\end{figure}
}{}%
}{}%

\begin{figure}[t]
  \centering
  \includegraphics[
    width=0.9\textwidth
  ]{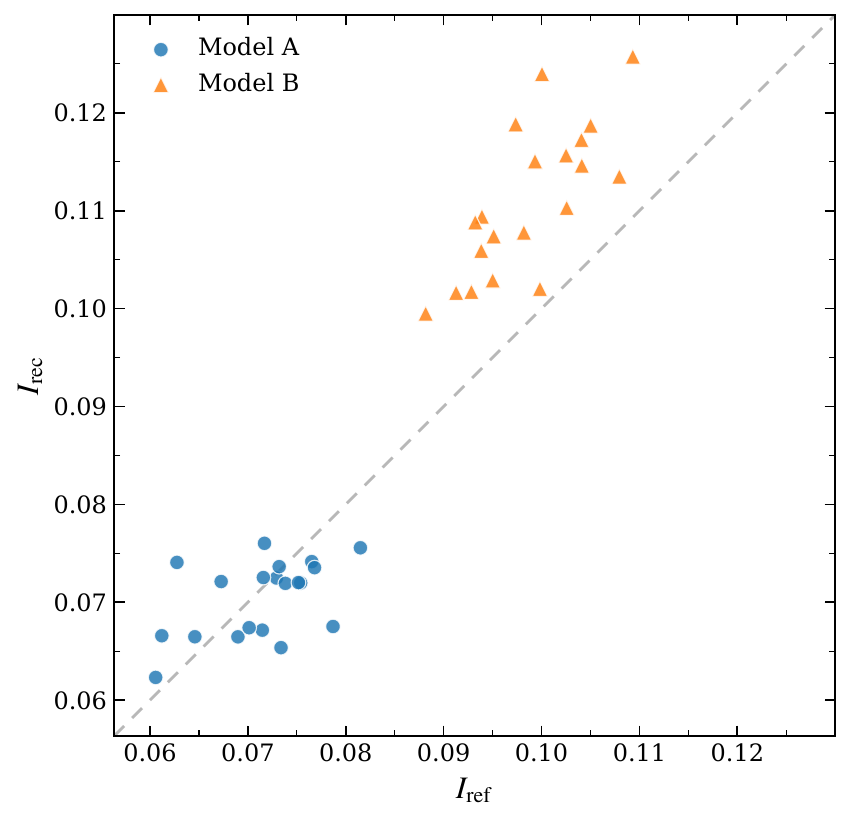}
  \caption{%
  Mass--redshift diagnostic for the two benchmark populations.
  Each point represents one training-run reconstruction and compares $I_{\rm rec}$ with the
  corresponding redshift-independent reference $I_{\rm ref}$, both conditioned
  on $\Stilde$.
  The dashed diagonal corresponds to $\DeltaIS=0$.
  Model~A lies close to and on both sides of the diagonal, whereas Model~B is
  displaced toward larger $I_{\rm rec}$ by the simulated evolution of the
  primary-mass peak.
  }
  \label{fig:benchmark-delta-I}
\end{figure}

\paragraph{Mass--redshift diagnostic.}
We next evaluate $\DeltaIS$ defined in
Eq.~\eqref{eq:global_evolution_diagnostic}.
To characterize the variation across training runs, we construct one density
for each of the twenty runs.
For run $r$, all checkpoints retained from that run are combined using their
within-run weights,
\begin{equation}
  q_r(\bm u)
  =
  \sum_{a\in\mathcal A_r}
  \omega_{a\mid r}\,q_{\phi_a}(\bm u),
  \label{eq:run_level_reconstruction}
\end{equation}
so that each training run contributes one reconstructed population.
The diagnostic is then evaluated separately for each $q_r$.
Figure~\ref{fig:benchmark-delta-I} shows the resulting
$(I_{\rm ref},I_{\rm rec})$ pairs.
The diagonal corresponds to $\DeltaIS=0$.

For Model~A the reconstructions lie close to and on both sides of the diagonal,
as expected for a population with no simulated mass evolution.
Model~B is instead systematically displaced toward larger $I_{\rm rec}$,
reflecting the simulated redshift evolution of the primary-mass peak.
The median values across the twenty reconstructions, together with their central
$90\%$ spread, are
\begin{equation}
  \Delta I_{\Stilde}^{\,{\rm A}}
  =
  -0.0021^{+0.0078}_{-0.0065},
  \qquad
  \Delta I_{\Stilde}^{\,{\rm B}}
  =
  0.0122^{+0.0097}_{-0.0070}.
  \label{eq:benchmark-delta-I-values}
\end{equation}
The stationary benchmark is therefore compatible with $\DeltaIS=0$, whereas
the evolving benchmark shows a clear displacement from the diagonal.

\FloatBarrier
\section{Application to GWTC-5.0}
\label{sec:gwtc5}

We now apply the pipeline to the binary-black-hole population in GWTC-5.0.
The reconstruction framework, support construction, ensemble weighting, and
mass--redshift diagnostic follow the procedure established and validated on the
benchmark simulations in Section~\ref{sec:pipeline-validation}.
Analysis-specific numerical settings are stated explicitly below.

\subsection{Catalogue and analysis inputs}
\label{sec:gwtc5-catalogue}

We analyse the binary-black-hole population in GWTC-5.0
\cite{GWTC5Catalogue}, using the Stable Release 6 event list
\cite{GWTC5CosmologyData}.
We retain binary-black-hole candidates detected with a false-alarm rate
\begin{equation}
  {\rm FAR}\le 0.25~{\rm yr}^{-1},
  \label{eq:gwtc5_far_cut}
\end{equation}
equivalently an inverse false-alarm rate of at least four years.
This leaves $C=231$ binary-black-hole events.

For each event we use the publicly released posterior samples specified by the
Stable Release 6 event list, together with the corresponding parameter-estimation
prior.\footnote{The corresponding posterior products are GWTC-2.1 v2 for O1,
O2, and O3a \cite{GWTC21PEData}, GWTC-3 v2 for O3b \cite{GWTC3PEData},
GWTC-4.0 for O4a \cite{GWTC4PEData}, and GWTC-5.0 Stable Release 6 for
O4b \cite{GWTC5SR6Part1,GWTC5SR6Part2}.}
We retain the detector-frame variables
$(D_L,m_{1,\rm det},m_{2,\rm det})$.

We do not infer the spin distribution, but adopt a fixed prescription with
uniform spin magnitudes and isotropic orientations, applied consistently in the
event and selection terms. The spin dependence of the injection proposal is
included in the selection weights.

\paragraph{Selection function.}
The selection correction is evaluated from the public cumulative search-sensitivity
injection campaign covering O1 through O4b
\cite{GWTC5CumulativeInjections}, in which
$N_{\rm tot}^{\rm gen}=1.57\times10^{9}$ injections were generated over an
observing time of $3.05$ yr.
An injection is counted as detected if it satisfies either the false-alarm-rate
threshold of Eq.~\eqref{eq:gwtc5_far_cut} or a semi-analytic network
signal-to-noise ratio of at least $10$.
Injections in the ER15 engineering interval are excluded, and both source-frame
component masses are required to exceed $3\,M_\odot$.
This mass cut applies to the injection set alone and is not imposed on the event
posterior samples.
The campaign leaves $N_{\rm det}^{\rm inj}=1.34\times10^{6}$ detected injections,
each carrying the proposal density $\pinj$ from which it was drawn.

\paragraph{Coordinates.}
Both the event samples and the injections are mapped from the detector-frame
variables to the source-frame parameters $\thv=(z,m_1,m_2)$ using a fixed
flat-$\Lambda$CDM cosmology, with
$H_0=67.9~{\rm km\,s^{-1}\,Mpc^{-1}}$ and $\Omega_m=0.3065$.
The cosmology is held fixed throughout: it defines the source-frame coordinates
and is not inferred.
The interim and injection priors are transformed with the Jacobian of this map and
of the subsequent standardization of Eq.~\eqref{eq:u_def_training}, so that every
importance ratio entering the hierarchical likelihood is evaluated with respect to
a common measure.
\subsection{Reconstructed population and mass--redshift dependence}
\label{sec:gwtc5-results}

We use the same overall reconstruction framework, support construction, ensemble
weighting, and diagnostic as for the benchmark catalogues.
For the observational analysis the event batch is set to $B=32$, so that the
fraction of the catalogue entering each update remains comparable to that of the
benchmark runs.
As in the benchmark analyses, $\Delta_{\rm cut}$ is chosen to retain checkpoints
from the near-optimal part of the validation-loss trajectory.
At the smaller event batch used for GWTC-5.0 the optimization trajectories show larger validation-loss fluctuations, so we adopt a wider gate,
$\Delta_{\rm cut}=0.06$ compared with $0.01$ for the benchmarks.

\paragraph{Reconstructed population.}
Figure~\ref{fig:gwtc5-z-bins} shows the reconstructed conditional primary-mass
distributions
$p(\log_{10}(m_1/M_\odot)\mid z\in Z_b,\Stilde)$
in fixed redshift intervals.
The reconstruction resolves two features in the primary-mass spectrum, near
$10\,M_\odot$ and near $35\,M_\odot$, the second accompanied by a change in the
slope of the distribution.
This structure is consistent with that inferred by the LVK population analysis of
the same catalogue \cite{GWTC5Population}.

\IfFileExists{figures/m1_z_slices_GWTC5_B32_ATF_F1B7_STYLE.pdf}{%
\begin{figure}[t]
  \centering
  \includegraphics[width=0.72\textwidth]{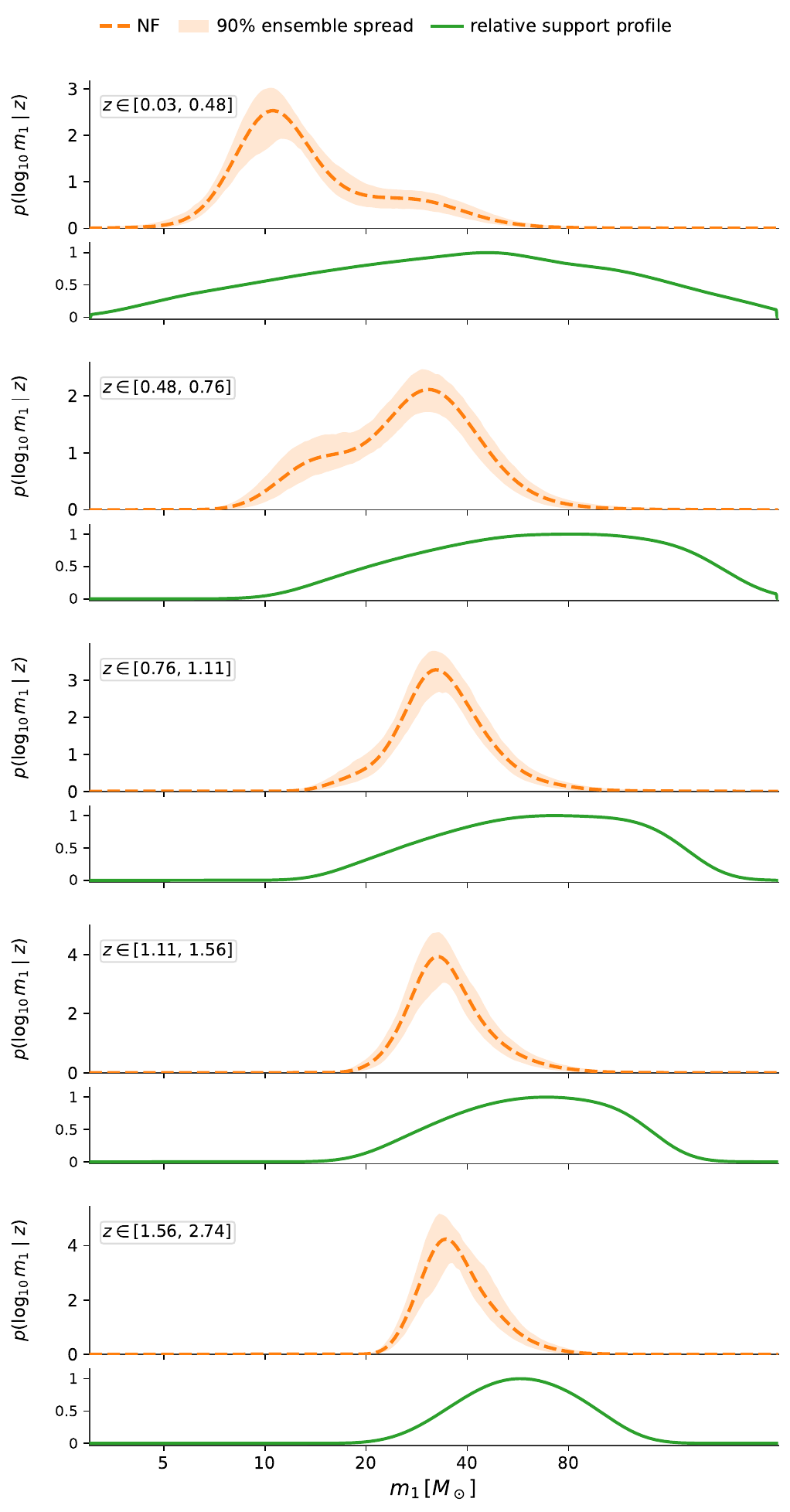}
  \caption{%
  Redshift-resolved primary-mass reconstruction for the GWTC-5.0 binary-black-hole
  sample, conditioned on the injection-supported region $\Stilde$.
  In each redshift bin the solid curve is the weighted ensemble mixture and the
  shaded band the central $90\%$ spread across retained members.
  The lower subpanels show the relative projected support volume as a function of
  primary mass, normalized within each redshift bin.
  }
  \label{fig:gwtc5-z-bins}
\end{figure}
}{}

\paragraph{Mass--redshift dependence.}
We evaluate $\DeltaIS$ of
Eq.~\eqref{eq:global_evolution_diagnostic} on the GWTC-5.0 reconstruction using
the same procedure as for the benchmarks.
The retained checkpoints are combined separately within each of the twenty
training runs according to Eq.~\eqref{eq:run_level_reconstruction}, and the
diagnostic is evaluated on each resulting reconstruction.
We obtain
\begin{equation}
  \Delta I_{\Stilde}^{\,\rm GWTC5}
  =
  0.0026^{+0.0049}_{-0.0043},
  \label{eq:gwtc5-delta-I}
\end{equation}
where the uncertainties denote the central $90\%$ spread across the
reconstructions.

As shown in Fig.~\ref{fig:gwtc5-delta-I}, the GWTC-5.0 reconstruction does not
show the clear positive displacement recovered for the evolving benchmark.
We therefore find no evidence for a dependence of the primary-mass spectrum on
redshift beyond that induced by the geometry of the injection-supported region.

This agrees with the non-parametric PixelPop reconstruction of the same catalogue
\cite{GWTC5Population} and with the parametric analysis of GWTC-3 in
Ref.~\cite{Lalleman2025NoMassEvolution}.
It differs from the non-parametric evolution reported for GWTC-3
\cite{Rinaldi2024Evolution} and the tentative high-mass evolution reported for
GWTC-4.0 \cite{AfrozMukherjee2025GWTC4MassEvolution}.

\IfFileExists{figures/Irec_vs_Iref_GWTC5.pdf}{%
\begin{figure}[t]
  \centering
  \includegraphics[width=0.9\textwidth]{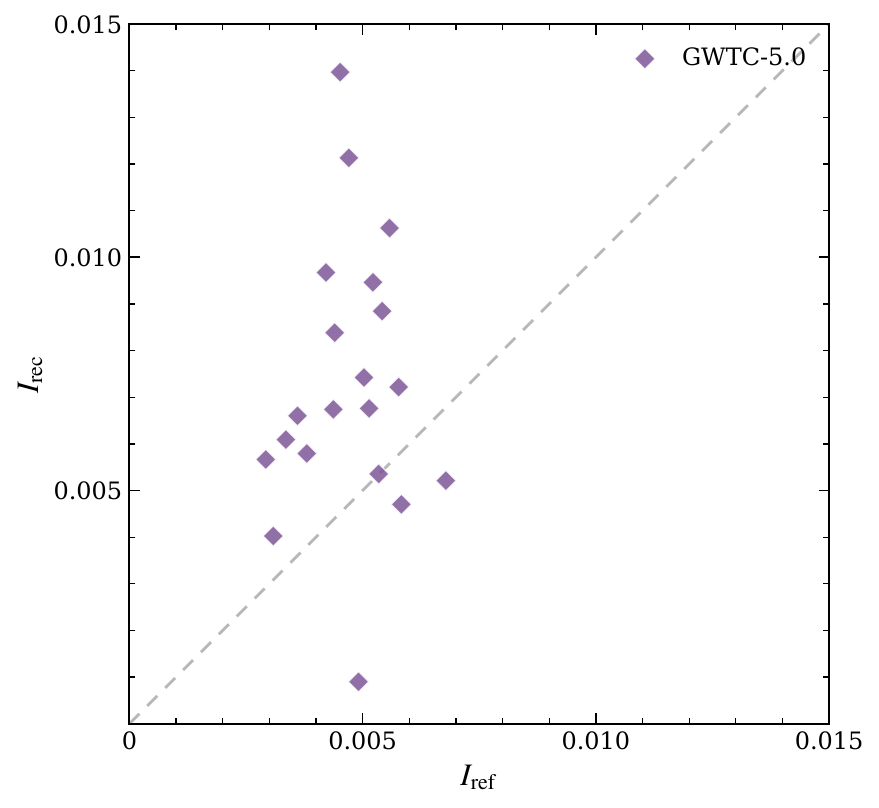}
  \caption{%
  Mass--redshift diagnostic for the GWTC-5.0 reconstruction.
  Each point represents one training-run reconstruction and compares
  $I_{\rm rec}$ with the corresponding redshift-independent reference
  $I_{\rm ref}$, both conditioned on $\Stilde$.
  The dashed diagonal corresponds to $\DeltaIS=0$.
  GWTC-5.0 does not show the clear positive displacement recovered for the
  evolving benchmark.
  }
  \label{fig:gwtc5-delta-I}
\end{figure}
}{}%

\FloatBarrier
\section{Conclusions}
\label{sec:conclusions}

We have presented a normalizing-flow framework for gravitational-wave population
inference in which the population density is optimized directly through the
hierarchical likelihood. The method reconstructs the joint source-frame
distribution in $(z,m_1,m_2)$ while incorporating event-level measurement
uncertainty and correcting for selection effects through detected injections.
It therefore provides an approach to multidimensional population reconstruction
without requiring a low-dimensional parametric mass model.

We validated the framework on two controlled binary-black-hole populations.
Model~A has a redshift-independent mass distribution, whereas Model~B contains an
evolving component in the primary-mass spectrum. In both cases the reconstruction
recovers the injected mass structure within the supported region. The
mass--redshift diagnostic is consistent with zero for Model~A and shows a clear
positive displacement for Model~B, thereby distinguishing the stationary and
evolving benchmarks.

The numerical tests show that the stochastic evaluation of the hierarchical
likelihood is well controlled at the adopted settings. Updates computed from
finite event batches remain closely aligned with the full-catalogue update, while
posterior-sample and injection subsampling contribute less variability than the
composition of the event batch. Together with the repeated training runs, these
tests show that the reconstructed benchmark features and their diagnostic
separation are not driven by the Monte Carlo approximations used during training.
The remaining differences between the true and reconstructed populations should
be interpreted in the context of finite catalogue size and of using a
maximum-likelihood reconstruction within a finite normalizing-flow family.

We then applied the framework to the GWTC-5.0 binary-black-hole catalogue.
The reconstruction recovers features consistent with the LVK population analysis,
including structure near $10\,M_\odot$ and $35\,M_\odot$, without imposing a
parametric mass model. The mass--redshift diagnostic does not show the clear
positive displacement recovered for the evolving benchmark. We therefore find no
evidence for a dependence of the primary-mass spectrum on redshift beyond that
induced by the geometry of the injection-supported region.

A natural extension is to include cosmological parameters directly in the
inference, allowing the source-frame population and the distance--redshift
relation to be constrained jointly. More broadly, the same statistical structure
arises whenever individual objects are measured with uncertainty and observed
through a process that preferentially selects some regions of parameter space.
Examples include exoplanet demographics, supernova surveys, and epidemiological
data with under-ascertainment. The combination of hierarchical likelihoods,
neural density models, and selection-aware training therefore provides a general
framework for reconstructing population distributions from incomplete and biased
observations.
\acknowledgments

The authors acknowledge the support of the computing facilities at INFN Rome of
the Amaldi Research center funded by the MIUR program ``Dipartimento di
Eccellenza'' (CUP: B81I18001170001).
This material is based upon work supported by NSF's LIGO Laboratory which is a
major facility fully funded by the National Science Foundation.

\clearpage
\appendix


\appendix

\section{Autoregressive neural spline flow construction}
\label{app:flow-details}

To construct a multivariate bijection in $D$ dimensions, we use an
\emph{autoregressive} flow \cite{papamakarios2017maf}. In an autoregressive transform,
each component is mapped using parameters that depend only on the preceding components
in a fixed ordering. Denoting the input to a flow step by $\vec u_{k-1}$ and its output
by $\vec u_k$, an autoregressive step can be written as
\begin{equation}
(\vec u_k)_d
=
T_d\!\left((\vec u_{k-1})_d;\, \eta_d\!\left((\vec u_{k-1})_{<d}\right)\right),
\qquad d=1,\ldots,D,
\label{eq:autoregressive_form}
\end{equation}
where $T_d(\cdot\,;\cdot)$ is an invertible, monotone one-dimensional transform acting on
the $d$th coordinate, and $\eta_d(\cdot)$ denotes the coordinate-specific output of a
conditioner network that takes the earlier coordinates $(\vec u_{k-1})_{<d}$ as input and
returns the parameters of $T_d$.

With this structure, the transformed coordinate $(\vec u_k)_d$ depends only on
$(\vec u_{k-1})_{\le d}$. As a result, the Jacobian of each flow step is triangular, so
the determinant entering Eq.~\eqref{eq:flow_changevar} reduces to the product of the
coordinate-wise derivatives
$\partial (\vec u_k)_d/\partial (\vec u_{k-1})_d$. This makes the density update at each
step straightforward to evaluate.

In our implementation, we choose $T_d$ in Eq.~\eqref{eq:autoregressive_form} to be a
monotonic rational-quadratic spline, yielding an \emph{autoregressive neural spline flow}
\cite{Durkan2019NeuralSpline}. A rational-quadratic spline defines a smooth, strictly
monotone map $x\mapsto y=s(x)$ on $\mathbb{R}$. The transformation $s$ is constructed on a
bounded interval $[-b_{\rm tail},b_{\rm tail}]$ by partitioning it into $L$ bins and defining a piecewise
rational-quadratic function on each bin. Outside $[-b_{\rm tail},b_{\rm tail}]$, a simple tail rule (commonly
linear) is used, ensuring that $s:\mathbb{R}\to\mathbb{R}$ remains a bijection.

The spline is specified by the knot locations $\{x_\ell\}_{\ell=0}^{L}$, the knot values
$\{y_\ell\}_{\ell=0}^{L}$, the positive derivatives $\{s'(x_\ell)\}_{\ell=0}^{L}$ at the
knots to enforce monotonicity, and the tail rule outside $[-b_{\rm tail},b_{\rm tail}]$.

In neural spline flows, these spline parameters are produced by the conditioner network and
constrained to satisfy the positivity and monotonicity requirements (e.g.\ via
softplus/softmax parameterizations) \cite{Durkan2019NeuralSpline}. This gives a flexible
family of one-dimensional monotone transformations while keeping both the inverse map and
the derivative needed for the Jacobian determinant analytically tractable.
\FloatBarrier

\section{Training details}
\label{app:training-details}

The settings below describe the two benchmark analyses; settings specific to the
GWTC-5.0 analysis are given in Section~\ref{sec:gwtc5-results}.
The main architecture and training settings are collected in
Table~\ref{tab:training-settings}.

\subsection{Flow architecture}
\label{app:training-architecture}

The population density is modelled by a stack of autoregressive rational-quadratic
spline transformations \cite{papamakarios2017maf,Durkan2019NeuralSpline}, initialized
to the identity map.
The variables are not permuted between transformations but keep the fixed ordering
$(z,m_1,m_2)$, giving the autoregressive factorization
\begin{equation}
  p(z,m_1,m_2)=p(z)\,p(m_1\mid z)\,p(m_2\mid z,m_1).
  \label{eq:autoregressive_factorization}
\end{equation}
This ordering follows the physical structure of the population, in which the masses
depend on redshift.
Spectral normalization \cite{Miyato2018SpectralNorm} is applied to the hidden linear
layers of the conditioner networks, which regularizes them and keeps the optimization
stable.
 
\subsection{Stochastic estimation of the objective}
\label{app:training-batching}

Each optimizer step computes the hierarchical loss using $B$ events, $K$ posterior
samples per selected event, and $M$ detected injections.
We justify the choice of these three parameters below.

\paragraph{Events per update.}
The event batch size controls how closely an individual stochastic update follows
the update implied by the full training catalogue.
We quantify this at two frozen states of a dedicated pilot optimization: one after
the learning-rate warmup and one late in training.
At each state, the posterior and injection Monte Carlo realizations are held fixed,
so that only the composition of the event batch varies.
They are evaluated using $K=8192$ posterior samples per event and
$M=3\times10^{5}$ injections, above the values used during training, to reduce
Monte Carlo fluctuations in this comparison.

For a batch of $B$ events, we compute the Adam update $\bm u_B$ using the same
optimizer state and gradient clipping as in training.
The reference update $\bm u_\star$ is obtained from all
$N_{\rm train}=1024$ training events.
Their directional agreement is measured by
\begin{equation}
  C_B^{\rm Adam}
  =
  \frac{
    \langle\bm u_B,\bm u_\star\rangle
  }{
    \lVert\bm u_B\rVert\,
    \lVert\bm u_\star\rVert
  },
  \label{eq:update_cosine}
\end{equation}
with $C_B^{\rm Adam}=1$ when the two update directions coincide.
This quantity compares the actual optimizer updates rather than the instantaneous
batch gradients; the stored Adam momentum is therefore common to all subsets at
a given frozen state.
For each batch size, Table~\ref{tab:batch-size-audit} reports the median
$C_B^{\rm Adam}$ and the $[5\%,95\%]$ interval over $256$ random event subsets.

\begin{table}[htbp]
  \centering
  \setlength{\tabcolsep}{14pt}
  \renewcommand{\arraystretch}{1.15}
  \begin{tabular}{@{}ccc@{}}
    \toprule
    $B$ & post-warmup & late training \\
    \midrule
    $32$  & $0.946\,[0.888,\,0.975]$ & $0.946\,[0.889,\,0.974]$ \\
    $64$  & $0.956\,[0.906,\,0.983]$ & $0.958\,[0.915,\,0.982]$ \\
    $128$ & $0.975\,[0.938,\,0.988]$ & $0.973\,[0.936,\,0.992]$ \\
    $256$ & $0.987\,[0.963,\,0.996]$ & $0.988\,[0.968,\,0.996]$ \\
    \bottomrule
  \end{tabular}
  \caption{%
  Alignment of finite-batch Adam updates with the update obtained from the full
  training catalogue. Entries give the median $C_B^{\rm Adam}$ and the
  $[5\%,95\%]$ interval over random event subsets.
  }
  \label{tab:batch-size-audit}
\end{table}

\noindent
The smaller batches $B=32$ and $64$ show appreciably weaker agreement with the
full-catalogue update.
At $B=128$, the median cosine is approximately $0.97$ at both stages of training,
showing that the stochastic update remains closely aligned with the
catalogue-level direction.
Increasing the batch to $B=256$ improves the alignment further, so the agreement
has not saturated at $B=128$ within the tested range.
We therefore retain $B=128$ for the benchmark analyses.
Each update then evaluates half as many event contributions as at $B=256$.
The audit characterizes the fidelity of this choice rather than identifying a
unique optimal batch size.

\paragraph{Posterior samples per event.}
Event $i$ contributes to the hierarchical loss through the log-mean-exp of its
importance-weighted posterior samples,
\begin{equation}
  \ell_i(K)
  =
  \log\!\left[
    \frac{1}{K}\sum_{k=1}^{K}
    \frac{\qphi(\bm u_{i,k})}{\pi_{i,u}(\bm u_{i,k})}
  \right],
  \label{eq:event_logmeanexp}
\end{equation}
which enters the objective with a negative sign.
A finite value of $K$ introduces both a small systematic offset in this
logarithmic estimator and stochastic variation between posterior subsamples.

We assess both contributions at the same late training state used in the
batch-size audit.
Writing $\ell_i^\star$ for the same quantity evaluated using all $32\,768$
available posterior samples, we draw $32$ independent finite-$K$ subsamples per
event.
The resulting mean bias and Monte Carlo scatter are reported in
Table~\ref{tab:posterior-sample-audit}.
The final column gives the standard deviation of the corresponding contribution
to the loss after averaging over an update containing $B=128$ events.

\begin{table}[htbp]
  \centering
  \setlength{\tabcolsep}{11pt}
  \renewcommand{\arraystretch}{1.15}
  \begin{tabular}{@{}cccc@{}}
    \toprule
    $K$
    & mean bias
    & MC scatter
    & $\mathrm{sd}$ at $B=128$ \\
    \midrule
    $512$
    & $-9.5\times10^{-4}$
    & $0.0443$
    & $3.94\times10^{-3}$ \\
    $1024$
    & $-6.2\times10^{-4}$
    & $0.0310$
    & $2.78\times10^{-3}$ \\
    $2048$
    & $-2.7\times10^{-4}$
    & $0.0222$
    & $1.98\times10^{-3}$ \\
    $4096$
    & $-1.5\times10^{-4}$
    & $0.0158$
    & $1.41\times10^{-3}$ \\
    $8192$
    & $-1.2\times10^{-4}$
    & $0.0110$
    & $0.98\times10^{-3}$ \\
    \bottomrule
  \end{tabular}
  \caption{%
  Finite-$K$ error in the posterior contribution to the hierarchical loss.
  The final column gives the stochastic variation after averaging over an
  update containing $B=128$ events.
  }
  \label{tab:posterior-sample-audit}
\end{table}

\noindent
At the value used in training, $K=1024$, the systematic offset in the per-event
contribution is below $10^{-3}$.
Although individual event estimates fluctuate more strongly, averaging over the
$B=128$ events in an update reduces the corresponding stochastic variation to
$2.8\times10^{-3}$.
Increasing to $K=2048$ reduces this value to $2.0\times10^{-3}$ while doubling
the number of posterior samples evaluated in the event term.
We therefore use $K=1024$ during training.
The reference $\ell_i^\star$ is itself a finite-sample estimate, so this audit
quantifies the error introduced by reducing $K$ below the available $32\,768$
posterior samples and not the residual Monte Carlo error of the reference itself.

\paragraph{Injections per update.}
The number of detected injections $M$ controls the Monte Carlo variation of the
selection term.
For each value of $M$ we evaluate $256$ independently drawn injection subsets and
quantify the relative variation of the resulting selection-factor estimates through
\begin{equation}
  \mathrm{CV}(\hat\alpha)
  =
  \frac{\mathrm{sd}(\hat\alpha)}
       {\mathbb E[\hat\alpha]}.
\end{equation}
Because the hierarchical objective contains $\log\hat\alpha$, we also report the
standard deviation of this logarithmic contribution.
The results are given in Table~\ref{tab:injection-sample-audit}.

\begin{table}[htbp]
  \centering
  \setlength{\tabcolsep}{14pt}
  \renewcommand{\arraystretch}{1.15}
  \begin{tabular}{@{}ccc@{}}
    \toprule
    $M$
    & selection-factor CV
    & $\mathrm{sd}(\log\hat\alpha)$ \\
    \midrule
    $5\times10^{4}$
    & $0.463\%$
    & $4.63\times10^{-3}$ \\
    $10^{5}$
    & $0.314\%$
    & $3.13\times10^{-3}$ \\
    $2\times10^{5}$
    & $0.224\%$
    & $2.24\times10^{-3}$ \\
    $3\times10^{5}$
    & $0.195\%$
    & $1.95\times10^{-3}$ \\
    $5\times10^{5}$
    & $0.155\%$
    & $1.55\times10^{-3}$ \\
    \bottomrule
  \end{tabular}
  \caption{%
  Monte Carlo variation of the selection term as a function of the number of
  detected injections used in each update.
  }
  \label{tab:injection-sample-audit}
\end{table}

\noindent
The measured variation follows the expected approximately $M^{-1/2}$ decrease.
At the value used in training, $M=10^{5}$, the selection-factor CV is only
$0.314\%$, with $\mathrm{sd}(\log\hat\alpha)=3.1\times10^{-3}$.
Larger injection subsets therefore reduce an already sub-percent fluctuation
while increasing the number of selection-function evaluations at every update.
We use $M=10^{5}$ during training.

\paragraph{Combined Monte Carlo check.}
As a joint check of posterior- and injection-sampling noise, we repeat the
full-catalogue gradient calculation using eight independent Monte Carlo
realizations at the training values $(K,M)=(1024,10^{5})$.
The resulting Adam updates have a median pairwise cosine of $0.9994$.
This is substantially closer to unity than the finite-event-batch values above,
showing that posterior- and injection-sampling noise is subdominant to the
variation associated with event-batch composition.

\subsection{Training campaign and checkpoint selection}
\label{app:training-runs}

The twenty training runs are organized as four independent five-fold partitions of the
catalogue.
Within each partition, every fold serves once as the validation set while the remaining
$80\%$ of the events are used for optimization.
Each run uses an independent random seed, so the ensemble samples both changes in the
training--validation partition and stochastic optimization.
Each run contributes equal total weight to the final ensemble.

Optimization uses Adam with gradient-norm clipping and a linear warmup to the
principal learning rate, after which the learning rate is reduced when the
validation loss plateaus. Training is terminated by early stopping based on the
validation loss.

Checkpoints are saved along each trajectory, with the cadence thinned as the learning
rate decreases so that the late, closely spaced and strongly correlated part of the trajectory
does not dominate the candidate pool; validation-best states are always kept.
Every saved checkpoint is then rescored under one common deterministic evaluation
on its run's held-out validation fold, the score being the mean per-event loss over
that fold so that runs remain comparable; this score alone governs selection and
weighting.
Checkpoints whose score lies within $\Delta_{\rm cut}$ of their run's best enter the
ensemble; the value of $\Delta_{\rm cut}$ is given in
Table~\ref{tab:training-settings}.

\begin{table}[t]
  \centering
  \begin{tabular}{@{}ll@{}}
    \toprule
    \multicolumn{2}{@{}l}{\textit{Flow architecture}}\\
    \quad Autoregressive spline blocks     & 12 \\
    \quad Spline bins                      & 16 \\
    \quad Hidden width                     & 96 \\
    \quad Residual blocks per conditioner  & 2 \\
    \quad Base distribution                & diagonal Gaussian \\
    \addlinespace[0.4em]

    \multicolumn{2}{@{}l}{\textit{Stochastic estimation}}\\
    \quad Events per update $B$            & 128 \\
    \quad Posterior samples per event $K$  & 1024 \\
    \quad Injections per update $M$        & $10^{5}$ \\
    \addlinespace[0.4em]

    \multicolumn{2}{@{}l}{\textit{Optimization}}\\
    \quad Optimizer                        & Adam \\
    \quad Principal learning rate          & $10^{-4}$ \\
    \quad Linear warmup                    & 15 epochs \\
    \quad Gradient-norm clip               & 5 \\
    \quad Learning-rate decay factor       & 0.5 \\
    \quad Learning-rate scheduler patience & 3 checks \\
    \quad Early-stopping patience          & 10 checks \\
    \addlinespace[0.4em]

    \multicolumn{2}{@{}l}{\textit{Cross-validation and ensemble}}\\
    \quad Folds per partition              & 5 \\
    \quad Independent partitions           & 4 \\
    \quad Checkpoint gate $\Delta_{\rm cut}$ & 0.01 \\
    \bottomrule
  \end{tabular}
  \caption{%
  Main architecture and training settings for the two benchmark populations.
  }
  \label{tab:training-settings}
\end{table}
\FloatBarrier

\section{Mock catalogues and injection campaigns}
\label{app:mock-catalogues}

Starting from the benchmark populations of Section~\ref{sec:benchmark-populations},
we construct for each model a mock detected catalogue, synthetic event-level
posterior samples, and an injection campaign.
The cosmology, measurement model, and detection rule are identical for Models~A
and~B.

\subsection{Detection model}
\label{app:mock-detection-model}

For a source with source-frame parameters $\thv=(z,m_1,m_2)$ we compute the
luminosity distance in the same flat-$\Lambda$CDM cosmology used to define the
benchmark populations,
\begin{equation}
  D_L(z)
  =
  \frac{c\,(1+z)}{H_0}
  \int_0^z
  \frac{\dd z'}{\sqrt{\Omega_m(1+z')^3+(1-\Omega_m)}},
  \label{eq:dl_of_z}
\end{equation}
where $c$ is the speed of light.
We define the detector-frame component masses
$m_{1,\rm det}=(1+z)m_1$ and $m_{2,\rm det}=(1+z)m_2$, the detector-frame total
mass $M_{\rm det}=m_{1,\rm det}+m_{2,\rm det}$, and the detector-frame chirp mass
$\mathcal M_{\rm det}=(m_{1,\rm det}m_{2,\rm det})^{3/5}
(m_{1,\rm det}+m_{2,\rm det})^{-1/5}$.
The true network signal-to-noise ratio is then
\begin{equation}
  \rho_{\rm true}
  =
  \rho_{\rm ref}\,\Theta\,
  \left(\frac{\mathcal M_{\rm det}}{\mathcal M_{\rm ref}}\right)^{5/6}
  \left(\frac{D_{L,\rm ref}}{D_L}\right)
  S\!\left[f_{\rm ISCO}(M_{\rm det})\right],
  \label{eq:snr_model}
\end{equation}
where $(\rho_{\rm ref},\mathcal M_{\rm ref},D_{L,\rm ref})
=(12,\,25\,M_\odot,\,2.5\,{\rm Gpc})$ set the reference scale of the unsuppressed
signal-to-noise ratio and $S$ applies an additional total-mass-dependent
suppression.
The orientation factor $\Theta$ is drawn by inverse-transform sampling from the
tabulated five-detector cumulative distribution used in the simulation,
restricted to $\Theta\in[\Theta_{\min},\Theta_{\max}]$ with
$(\Theta_{\min},\Theta_{\max})=(0.2,1.6)$.

The suppression factor is built from the innermost-stable-circular-orbit
frequency,
\begin{equation}
  f_{\rm ISCO}(M_{\rm det})
  =
  \frac{1}{\pi\,6^{3/2}\,M_{\rm det}\,M_\odot^{\rm sec}},
  \qquad
  M_\odot^{\rm sec}=5\times10^{-6}\,{\rm s},
  \label{eq:fisco}
\end{equation}
through the logistic transition
\begin{equation}
  S(f)
  =
  \left[1+\exp\!\left(-k(f-f_{\rm mid})\right)\right]^{-1},
  \qquad
  f_{\rm mid}=\tfrac12(f_{\rm low}+f_{\rm high}),
  \qquad
  k=\frac{\ln(10^3)}{f_{\rm high}-f_{\rm low}},
  \label{eq:snr_suppression}
\end{equation}
with $(f_{\rm low},f_{\rm high})=(20,100)\,{\rm Hz}$.

To emulate measurement fluctuations for a network of $N_{\rm det}$ detectors we
draw an observed signal-to-noise ratio from a noncentral $\chi^2$ model,
\begin{equation}
  \rho_{\rm obs}=\sqrt{X},
  \qquad
  X\sim\chi'^2\!\left(2N_{\rm det},\,\rho_{\rm true}^2\right),
  \label{eq:snr_obs}
\end{equation}
with $N_{\rm det}=5$, and classify a source as detected when
$\rho_{\rm obs}\ge\rho_\star$ with $\rho_\star=12$.
The same rule is applied to the population draws forming the event catalogues and
to the injections.
Each catalogue contains $C=1280$ detected events.

\subsection{Synthetic event-level posterior samples}
\label{app:mock-posterior-samples}

For each detected event we generate posterior samples in the detector-frame
variables $(\rho^2,\ln\mathcal M_{\rm det},q,\Theta)$, where
$q=m_{2,\rm det}/m_{1,\rm det}$ is the mass ratio, related to the symmetric mass
ratio by $\eta(q)=q/(1+q)^2$.
We first construct noisy point estimates
$(\rho_{\rm obs},\ln\mathcal M_{\rm det,obs},\eta_{\rm obs},\Theta_{\rm obs})$:
the observed signal-to-noise ratio is drawn from Eq.~\eqref{eq:snr_obs}, and,
conditional on it, the remaining estimates are obtained by adding Gaussian noise
to the true detector-frame values with standard deviations that scale as
$1/\rho_{\rm obs}$,
\begin{equation}
  \sigma_{\ln\mathcal M}=0.08\,\frac{8}{\rho_{\rm obs}},
  \qquad
  \sigma_\eta=0.022\,\frac{8}{\rho_{\rm obs}},
  \qquad
  \sigma_\Theta=0.21\,\frac{8}{\rho_{\rm obs}}.
  \label{eq:pe_noise_model}
\end{equation}

Posterior samples are then drawn around these point estimates.
For the signal-to-noise-ratio coordinate we adopt the synthetic posterior density
\begin{equation}
  p(\rho^2\mid\rho_{\rm obs})
  \propto
  f_{\chi'^2}\!\left(\rho^2;\,2N_{\rm det},\,\rho_{\rm obs}^2\right)
  \left(\rho^2\right)^{-(N_{\rm det}-1)},
  \label{eq:rho_posterior_model}
\end{equation}
where $f_{\chi'^2}(x;\nu,\lambda)$ is the noncentral-$\chi^2$ density with $\nu$
degrees of freedom and noncentrality parameter $\lambda$.
The chirp mass and orientation factor are drawn from
\begin{equation}
  \ln\mathcal M_{\rm det}
  \sim
  \mathcal N\!\left(\ln\mathcal M_{\rm det,obs},\sigma_{\ln\mathcal M}^2\right),
  \qquad
  \Theta
  \sim
  \mathcal N\!\left(\Theta_{\rm obs},\sigma_\Theta^2\right),
\end{equation}
with $\Theta$ truncated to $[\Theta_{\min},\Theta_{\max}]$.
For the mass ratio the likelihood is Gaussian in $\eta(q)$ with width
$\sigma_\eta$, and we adopt an interim prior that is flat in $q$ on
$[q_{\min},q_{\max}]=[10^{-4},1]$, so that the mass-ratio samples follow
\begin{equation}
  p(q\mid\eta_{\rm obs})
  \propto
  \exp\!\left[
    -\frac{\left(\eta(q)-\eta_{\rm obs}\right)^2}{2\sigma_\eta^2}
  \right]
  \mathbb I\!\left(q_{\min}\le q\le q_{\max}\right).
  \label{eq:q_flat_prior}
\end{equation}

Each event is represented by $K_i=32\,768$ posterior samples in the detector-frame
variables $\thv_{\rm det}=(D_L,m_{1,\rm det},m_{2,\rm det})$, obtained from
$(\mathcal M_{\rm det},q)$ and by inverting Eq.~\eqref{eq:snr_model} for $D_L$.
The samples carry an approximately flat interim prior in $\thv_{\rm det}$, and
this is the interim prior $\pi_i$ entering the hierarchical likelihood.

\subsection{Injection campaign}
\label{app:mock-injections}

The injection prior is defined directly in detector-frame component masses and
luminosity distance.
Primary masses are drawn from a truncated power law
$\pinj(m_{1,\rm det})\propto m_{1,\rm det}^{-2}$ on
$m_{1,\rm det}\in[3.5,300]\,M_\odot$; conditional on the primary mass, secondary
masses are drawn from
$\pinj(m_{2,\rm det}\mid m_{1,\rm det})\propto m_{2,\rm det}$ on
$m_{2,\rm det}\in[3.5\,M_\odot,\,m_{1,\rm det}]$; and the luminosity distance is
drawn from
\begin{equation}
  \pinj(D_L)
  =
  \frac{3D_L^2}{D_{L,\max}^3-D_{L,\min}^3},
  \qquad
  D_L\in[D_{L,\min},D_{L,\max}],
  \label{eq:inj_dL_prior}
\end{equation}
with $(D_{L,\min},D_{L,\max})=(0,8)\,{\rm Gpc}$.

For every generated injection we evaluate $\rho_{\rm true}$ and $\rho_{\rm obs}$
from Eqs.~\eqref{eq:snr_model}--\eqref{eq:snr_obs} and apply the same detection
rule $\rho_{\rm obs}\ge\rho_\star$ used for the events.
Each campaign is continued until approximately $10^{6}$ detected injections have
accumulated.
For the detected subset $\{\thv_j^{\rm inj}\}_{j=1}^{N_{\rm det}^{\rm inj}}$ we
store the injection-prior density $\pinj(\thv_j^{\rm inj})$ and record the total
number of generated injections $N_{\rm tot}^{\rm gen}$; both enter the Monte Carlo
estimator of the selection factor, Eq.~\eqref{eq:alpha_mc}.
\FloatBarrier

\section{Construction of the injection-supported region}
\label{app:support}

The hierarchical selection correction is evaluated from the detected injections
through Eq.~\eqref{eq:alpha_mc}, and the same injection campaign delimits the
region of parameter space in which that correction is supported.
A detected injection is a draw from
\begin{equation}
  \pi(\thv\mid{\rm det})
  \propto
  \pdet(\thv)\,\pinj(\thv),
  \label{eq:detected_injection_density}
\end{equation}
whose support is the region $\mathcal S$ of Eq.~\eqref{eq:support_def}.
Since $\mathcal S$ is accessible only through a finite sample, we construct, for
each random split $\ell$ of the detected injections, an empirical acceptance
region $\Stilde_\ell$ with the property that a new detected draw falls inside it
with high probability,
\begin{equation}
  \mathbb P\!\left(\thv\in\Stilde_\ell\right)\ge 1-\alpha,
  \label{eq:coverage_app}
\end{equation}
for a chosen miscoverage level $\alpha$.
This is achieved by split-conformal prediction, which calibrates each region so
that Eq.~\eqref{eq:coverage_app} holds under the usual exchangeability
assumption.

We work in the standardized coordinates $\bm u(\thv)$ of
Eq.~\eqref{eq:u_def_training} and take as anomaly score the
$k$-nearest-neighbour radius
\begin{equation}
  s(\bm u)\equiv r_k(\bm u),
  \label{eq:knn_score}
\end{equation}
the Euclidean distance from $\bm u$ to its $k$-th nearest neighbour among a
reference set of detected injections.
A point far from the detected-injection cloud has a large $r_k$, so thresholding
the score isolates the region the injections populate.

The threshold is calibrated by splitting the detected injections into a training
set $D_{\rm tr}$, which defines the score, and a calibration set $D_{\rm cal}$,
which fixes the threshold; we use an $80/20$ split.
Sorting the calibration scores $s_j=s(\bm u_j)$, $\bm u_j\in D_{\rm cal}$, into
order statistics $s_{(1)}\le\cdots\le s_{(m)}$ with $m=|D_{\rm cal}|$, the
conformal quantile is
\begin{equation}
  t=s_{(h)},
  \qquad
  h=\left\lceil(1-\alpha)(m+1)\right\rceil,
  \label{eq:conformal_quantile}
\end{equation}
and the accepted region
$\Stilde_\ell=\{\thv:s^{(\ell)}[\bm u(\thv)]\le t_\ell\}$ then satisfies
Eq.~\eqref{eq:coverage_app} for a new detected draw.

A region constructed from one split depends on the particular
training--calibration partition of the detected injections.
We therefore repeat the construction for $n_{\rm split}=3$ separately randomized
$80/20$ partitions at the same $\alpha$ and retain their intersection,
\begin{equation}
  \Stilde=\bigcap_{\ell=1}^{n_{\rm split}}\Stilde_\ell.
  \label{eq:support_intersection}
\end{equation}
A point is then retained only if every partition accepts it, so points admitted
only by a single permissive split do not enter the final support estimate.

Each split has marginal miscoverage at most $\alpha$, and the intersection rejects
a point whenever any split does, so the union bound gives
\begin{equation}
  \mathbb P\!\left(\thv\in\Stilde\right)\ge 1-n_{\rm split}\alpha.
  \label{eq:coverage_intersection}
\end{equation}
We use $k=10$, $\alpha=0.01$ and $n_{\rm split}=3$, corresponding to a marginal
coverage of at least $0.97$ under the split-conformal exchangeability assumption.

All reported population summaries are obtained by conditioning the reconstruction on the
same region $\Stilde$.
\FloatBarrier

\section{Monte Carlo diagnostics}
\label{app:monte-carlo-diagnostics}

The hierarchical loss rests on two importance-sampling estimates, one over the
posterior samples of each event and one over the detected injections.
Either can be dominated by a few samples with large weights, so we monitor the
effective sample size of both.
For event $i$,
\begin{equation}
  N_{{\rm eff},i}^{\rm PE}
  =
  \frac{\left(\sum_k w_{i,k}\right)^2}{\sum_k w_{i,k}^2},
  \qquad
  w_{i,k}
  =
  \frac{\qphi(\bm u_{i,k})}{\pi_{i,u}(\bm u_{i,k})},
  \label{eq:ess_event}
\end{equation}
and for the selection integral,
\begin{equation}
  N_{\rm eff}^{\rm inj}
  =
  \frac{\left(\sum_j s_j\right)^2}
       {\displaystyle
        \sum_j s_j^2
        -\frac{1}{N_{\rm tot}^{\rm gen}}\left(\sum_j s_j\right)^2},
  \qquad
  s_j
  =
  \frac{\qphi(\bm u_j^{\rm inj})}{\pi_{{\rm inj},u}(\bm u_j^{\rm inj})}.
  \label{eq:neff_inj}
\end{equation}

Following the numerical-stability criteria commonly applied in gravitational-wave
population analyses \cite{Mastrogiovanni2023IcaroGW,TalbotGolomb2023LikelihoodUncertainty},
we compare these quantities with
\begin{equation}
  \min_i N_{{\rm eff},i}^{\rm PE}\ge 20,
  \qquad
  N_{\rm eff}^{\rm inj}>4C,
  \label{eq:mc_thresholds}
\end{equation}
with $C$ the number of catalogue events.
Both inequalities are satisfied by every reconstruction retained in this work.
\FloatBarrier

\clearpage

\IfFileExists{references.bib}{%
  \bibliographystyle{JHEP}
  \bibliography{references}
}{%
  \typeout{*** WARNING: `references.bib' not found; bibliography skipped. ***}%
}

\end{document}